\documentclass[conference]{IEEEtran}
\usepackage{amsmath}
\usepackage{amsthm}
\usepackage{amssymb}
\usepackage{verbatim}
\usepackage{graphicx}
\usepackage{tikz}
\usepackage{pgfplots}
\usetikzlibrary {positioning}
\usetikzlibrary{shapes,snakes}
\usetikzlibrary{arrows,calc}
\usepackage{relsize}
\usetikzlibrary{patterns}
 \usepackage{algorithm}
\usepackage{algpseudocode}

\theoremstyle{plain}
\newtheorem{lemma}{Lemma}
\newtheorem{rem}{Remark}
\newtheorem{theorem}{Theorem}

\newtheorem{proposition}[theorem]{Proposition}
\newtheorem{definition}{Definition}

\theoremstyle{definition}
\newtheorem{example}{Example}
\floatname{algorithm}{Algorithm}

\allowdisplaybreaks[4]

\usepackage{amsfonts}
\usepackage{times}
\usepackage{latexsym}
\usepackage{amssymb}
\usepackage{amsmath}
\usepackage{cite}
\usepackage{verbatim}
\usepackage{bbm}
\usepackage{tikz}
\usetikzlibrary{positioning,chains,fit,shapes,calc}
\usepackage[justification=centering]{caption}
\usepackage{graphicx,subcaption}
\usepackage{fancyhdr}
\usepackage{xcolor}

\usepackage{mathrsfs}
\usepackage{pgfplots}
\pgfplotsset{compat=1.14}
\tikzstyle{bipartite}=[thick,
  leftnode/.style={fill=blue,circle,draw,inner sep=0pt,minimum size=2mm},
  rightnode/.style={fill=green,circle,draw,inner sep=0pt,minimum size=2mm},
  dummynode/.style={fill=gray,circle,draw,inner sep=0pt,minimum size=2mm},
  ->,shorten >= 3pt,shorten <= 3pt
]
\newenvironment{customlegend}[1][]{%
    \begingroup
    \csname pgfplots@init@cleared@structures\endcsname
    \pgfplotsset{#1}%
}{%
    \csname pgfplots@createlegend\endcsname
    \endgroup
}%
\def\addlegendimage{\csname pgfplots@addlegendimage\endcsname}

\pgfkeys{/pgfplots/number in legend/.style={%
        /pgfplots/legend image code/.code={%
            \node at (0.295,-0.0225){#1};
        },%
    },
}

\tikzstyle{centerlabel}=[fill=white,
anchor=center,
midway]

\newtheorem{problem}{Problem}

\def\bb0{{\mathbb{0}}}

\def\bb{{\mathbf{b}}}

\def\b0{{\mathbf{0}}}
\def\opt{\mathsf{OPT}}

\def\b1{{\mathbf{1}}}

\def\bbE{{\mathbb{E}}}

\def\bbN{{\mathbb{N}}}

\def\cP{\mathcal{P}}

\def\sfO{\mathsf{O}}

\def\sfb{{\mathsf{b}}}

\def\sfr{{\mathsf{r}}}
\def\sfs{{\mathsf{s}}}

\def\sf0{{\mathsf{0}}}

\def\nn{\nonumber}

\begin{document}





\title{Not Just Age but Age and Quality of Information}

%
%

\author{\IEEEauthorblockN{Nived Rajaraman}
\IEEEauthorblockA{Indian Institute of Technology  Madras\\
 Mumbai, India\\
\textsf{nived.rajaraman@gmail.com}}
\and
\IEEEauthorblockN{Rahul Vaze}
\IEEEauthorblockN{
Tata Institute of Fundamental Research\\
  Mumbai \\
\textsf{rahul.vaze@gmail.com}}
\and 
\IEEEauthorblockN{Goonwanth Reddy}
\IEEEauthorblockA{University of Maryland\\
 College Park\\
\textsf{goonwanth.reddy@gmail.com}}
}

\maketitle

\begin{abstract}
A versatile scheduling problem to model a three-way tradeoff between delay/age, distortion, and energy is considered. The considered problem called the age and quality of information (AQI) is to select which packets to transmit at each time slot to minimize a linear combination of the distortion cost, the age/delay cost and the energy transmission cost in an online fashion.
AQI generalizes multiple important problems such as age of information (AoI), the remote estimation problem with sampling constraint,  
the classical speed scaling problem among others. The worst case input model is considered, where the performance metric is the competitive ratio. A greedy algorithm is proposed that is shown to be 2-competitive,  independent of all parameters of the problem. For the special case of AQI problem, a greedy online maximum weight matching based algorithm is also shown to be 2-competitive.
\end{abstract}

\section{Introduction} \label{sec:intro}
Consider multiple sources evolving over time with the $i^{th}$ sources' sample $s_i(t)$ arriving at a scheduler consisting of $B_i(t)$ bits at time $t$, e.g. a cyber-physical system such as multiple sensors in a car. 
From a quality of service (QoS) view, each of the 
sources would like to send as many of their bits to the receiver/monitor to have the smallest distortion, and as quickly as possible, to minimize delay/age, but there is a limit on the speed of transmission (accounted using an energy cost).

Thus, a three-fold tradeoff emerges between the information freshness (delay/age), the number of bits sent for each source (that controls distortion), and the total energy consumed. To model this tradeoff, we consider an objective function that is a linear combination of delay/age, distortion, and energy, which we call as age and quality of information (AQI) problem. 

In particular, sample $s_i(t)$ that arrives at time $t$ with $B_i(t)$ bits is divided into $\ell_i(t)$ equal sized sub-samples of size $\sfb$ (called sub-packets). Among the 
$\ell_i(t)$ sub-packets if $|S_p(t)| \le \ell_i(t)$ sub-packets are actually sent by the algorithm, where the last sub-packet among the $|S_p(t)|$ is sent at time $d_p(t)$, then the distortion cost is given by $D(|S_p(t)|)$ while the delay/age cost is $C_p(d_p(t)-t)$. Function $D$ is assumed to be sub-modular to capture the diminishing returns property, while $C$ is assumed to be convex. Moreover, if $k$ sub-packets (possibly belonging to different samples) are sent in time slot $t$, then energy cost is assumed to be $g(k)$ where $g(.)$ is a convex function. The overall objective function is a linear combination of $D(.), C_p(.)$ and $g(.)$ and the decision variable at each time slot is to send how many sub-packets among the outstanding ones.


AQI problem generalizes the age of information (AoI) problem, where the metric is the freshness of information at the receiver side, that has become a very popular object of theoretical interest in recent past \cite{kaul2012real, huang2015optimizing, sun2017update, yates2016age, najm2018content}. A nice review can be found in \cite{kosta2017age}.  
One important limitation of the AoI metric is that each source sample is binary, $B_i(t)=1$, i.e.,
the receiver is only interested in knowing whether a certain event has happened or not, and the objective is to minimize its staleness. 

AQI problem clearly has applications in transmission of video files, where files are coded in multiple resolutions/formats, HD or SD etc., and the objectives are to transmit files in as high a resolution possible subject to strict delay constraints and energy usage. Lot of  work has been accomplished in this direction \cite{zhou2010distributed, chakareski2006rate, pahalawatta2007content}, however, to the best of our knowledge not for the formulation of this paper. 

AQI problem was first considered in \cite{vaze2017energy} but only for the offline case, where a fixed number $n$ of packets (samples) with bits $B_i, 1\le i\le n$ are available at the transmitter at the start of communication. Some structural results for the optimal offline solution for the AQI problem were derived in \cite{vaze2017energy}, e.g., that the optimal order of transmission of packets is in increasing order of $B_i$, and that the offline AQI problem is jointly convex. The results of \cite{vaze2017energy} do not apply for the online setting, the focus of this paper. 

AQI problem is similar to the classical rate-distortion (RD) problem \cite{Cover2004}, where the objective is to find minimum transmission rate to support a given distortion constraint under a specific distortion metric. Typically, the RD problem is a considered for single source-destination pair, with infinitely large blocklengths \cite{Cover2004} or finite blocklengths \cite{gupta2008rate} and \cite{kostina2012fixed}. AQI problem can be seen as a slotted resource allocation analogue of the RD problem with multiple sources.


Many variants of the AoI problem have been considered in prior work \cite{yates2016age, huang2015optimizing, najm2018content}, with two main variants that are `near' special cases of the AQI problem are \cite{kadotascheduling} and \cite{DBLP:journals/corr/SunPU17}. In \cite{kadotascheduling}, multiple sources are considered, and at each time slot one bit of information from only one of the source (say $i$) can be communicated, and the objective is to minimize the long-term weighted sum of the ages of all sources, subject to individual source throughput constraints. An algorithm that is at most $2$ times the optimal cost is derived \cite{kadotascheduling}. The probabilistic communication model is also incorporated in \cite{kadotascheduling}, where each source's communication is successful with probability $p_i$. This problem generalizes the throughput maximization problem \cite{hou2009theory} for the AoI metric.

In \cite{DBLP:journals/corr/SunPU17}, a single source following a Wiener process is sampled at discrete time epochs, and the samples are sent to the receiver over a random delay channel under a first-come-first-come schedule (FCFS). The problem is to find the optimal sampling epochs so as to minimize the mean square error at the receiver under a sampling frequency constraint at the transmitter. 
This sampling problem \cite{DBLP:journals/corr/SunPU17} has elements that are similar to the AQI problem, like accounting for distortion because of infrequent sampling and random delay channel, however, it is limited to a single source and does not account for energy needed for transmitting the samples and the delay is not dependent on the size of the samples, since transmitting more bits for one sample delays the transmission for other samples.

 
The AQI problem can inherently capture the individual throughput constraints of each source (like in \cite{kadotascheduling}) by weighing the source distortion/delay appropriately in the objective function without explicitly enforcing it. 
Similarly, the sampling frequency constraint as considered in \cite{DBLP:journals/corr/SunPU17} for each source is implicitly included in the AQI problem without explicitly enforcing it, since there are multiple sources and the energy function is convex,
resulting in each source getting its transmission turn only at a certain rate similar to \cite{DBLP:journals/corr/SunPU17}. 

In this paper, we consider the worst case input setting for the AQI problem to model the most general setting, where each source sample $s_i(t)$ (arrival time or $B_i(t)$) need not follow any distribution and can even be chosen by an adversary. Under this worst case setting, the performance metric is the competitive ratio,  the ratio of the cost of any online algorithm and the optimal offline algorithm that knows all the future information, and the goal is to derive an online algorithm with least competitive ratio over all possible input sequences. Even though this setting appears too pessimistic, surprisingly we are able to derive algorithms that have competitive ratios of $2$, as detailed later.

AQI problem is also closely related to the 
classical speed scaling problem \cite{yao1995scheduling, bansal2009speedconf, wierman2009power, gupta2010scalably, gupta2012scheduling}, 
where jobs arrive over time and on its arrival each job has to be assigned to one of the multiple servers, where the server speed is variable, and running at speed $s$ incurs a cost of $P(s)$. 
Typically, the objective is to minimize a linear combination of the flow time and total energy consumption, where flow time is the sum of response time (departure-arrival time for any jobs) of all jobs. Near-optimal algorithms are known for speed scaling for a single server \cite{bansal2009speedconf, wierman2009power}, even in the worst case, but not for multiple servers, where speed augmentation is shown to be necessary (the online algorithm is allowed an extra speed of $1+\epsilon$ over the offline optimal algorithm) and the competitive ratio guarantees are $O(1/\epsilon)$ \cite{gupta2010scalably, gupta2012scheduling}. For the special case when $P(s) = s^\alpha$, the competitive ratio guarantees of $O(\alpha)$ are possible \cite{gupta2010scalably, gupta2012scheduling} without any speed augmentation, where the algorithms are a variant of processor sharing. 

The AQI problem considered in this paper can model a slotted version of the speed scaling problem with multiple servers, where the speed once chosen is fixed for a time slot, while in
the usual speed scaling problem speed can be changed continuously. 
AQI problem importantly enjoys an extra 'parallelizing' flexibility compared to the speed scaling problem, where different segments of each job can be processed on different servers. This feature of job splitting and parallel processing, is however, common in modern systems \cite{nelson1988performance, xing2000parallel}, e.g. mapreduce \cite{dean2008mapreduce}, and also inherently part of the processor sharing based algorithms \cite{gupta2010scalably, gupta2012scheduling} (best known algorithm for speed scaling with multiple servers), where in a short time span multiple jobs receive service from multiple servers.

\subsection{Contributions}
\begin{itemize}
\item We begin with a special case, the binary-AQI problem, where all samples $s_i(t)$ are of same size $B_i(t)=\sfb$, i.e., each packet has only one sub-packet, and the only decision choice is whether to transmit the sub-packet at all (in which time slot) or not at all.

This binary case is a generalization of the usual AoI metric \cite{kaul2012real, kadotascheduling} with an additional distortion and energy cost. 
For each time slot $t$, we define energy mini-slots $c_{kt}$, where $k$ represents the incremental  energy required to send an extra sub-packet of $\sfb$ bits in the same slot, i.e. $g'(k) - g'(k-1)$, where $g'(k) = g(kb)$, and $g(x)$ is the convex energy cost to send $x$ bits in a single time slot.
For the binary AQI problem, at time $t$, we define a bipartite graph between the outstanding sub-packets $\sfO(t)$ (left side nodes) and energy mini-slots $c_{kt'}, t'\ge t$ (right side nodes), where the edge weight between a source sample $s \in O(t)$ and mini-slot $c_{kt'}$ is 
$D(\sfb) - C(t'-t) - g'(k)-g'(k-1)$.

For this bipartite graph, we propose a local greedy online matching algorithm, that at each time slot $t$, finds a maximum weight matching using the causally available information, and at time slot $t$ transmits all the sub-packets that are matched to the mini-slots $c_{kt}$ of slot $t$, and progresses similarly for all time slots.

We show that this algorithm is $2$-competitive, thus finding a near optimal solution for the binary AQI even under the worst case. Recall that in \cite{kadotascheduling}, a  $2$-competitive algorithm has been 
derived only for AoI metric without accounting for distortion and energy cost, and moreover the results are in expectation and not for worst case. 

It is worth noting that no online maximum weight matching algorithm has a bounded competitive ratio when the edge weights are arbitrary \cite{secretary, korulapal, kesselheim}. To derive online algorithms with bounded competitive ratio \cite{korulapal, kesselheim} a secretarial model of input is assumed, where left side nodes arrive in a uniformly random order. We avoid this restriction by exploiting the exact utility function that governs the edge weights, (the edge weights on all outgoing edges for a packet are related) and show that a greedy algorithm is $2$-competitive. 

\item For the general AQI problem, each packet having multiple sub-packets makes the counting of delay/age harder, and we use a greedy algorithm to associate each outstanding sub-packet to the energy mini-slots (defined for the binary AQI problem), that provides the largest incremental increase in the AQI objective function. The general AQI problem is related to the standard sub-modular function maximization \cite{Fisher1978}, for which a greedy algorithm is known to have a competitive ratio of $2$.  The AQI problem  has several extra attributes/constraints compared to the standard sub-modular function maximization (see Problem \ref{RA:locking} definition),  and hence the above classical result does not hold directly for the AQI problem. 
We show that the proposed algorithm is also $2$-competitive for the general AQI problem, via carefully constructed reductions from the greedy algorithms for the standard sub-modular function maximization \cite{Fisher1978}, 
\end{itemize}


\section{System Model} 
We consider a slotted time system, and without loss of generality assume that each slot is of unit length. 
Consider multiple sources evolving over time with the $i^{th}$ sources' sample/packet $p_i(t)$ arriving at the scheduler at time slot $t$ consisting of $B_i(t)$ bits. To keep the exposition simple, we define costs for each packet corresponding to the same source separately. 
Source based costs can also be incorporated (as shown in Section \ref{subsec:aoi}), however, makes the notation cumbersome in general. Thus, in general, a set of packets $\cP(t)$ arrive at the scheduler at time slot $t$ from multiple sources, and the arrival time of packet $p$ is denoted by $A_p$.

Packet $p \in \cP(t)$ has $B_p$ bits which without loss of generality is assumed to be some multiple of $\sfb$ bits ($\sfb$ is a constant). We split each packet $p$ into $k_p = B_p/\sfb$ sub-packets, and define the cost of packet $p$ as follows. 

To keep the model most general, we define the utility function for packet $p \in \cP(A_p)$ to be time dependent as $V_p(t-A_p) : \mathbb{\bbN} \to \mathbb{R}$. For each packet $p$, let $1, \dots, |S_{p}(t)|$ denote the ordered sub-packets scheduled to be transmitted among its $k_p$ sub-packets by time $t$ (details in Remark \ref{rem:indexing}). Let $d_p(t)$ be the last time slot at which any of the $|S_{p}(t)|$ sub-packets are transmitted.
A simple but general enough example for $V_p$ at time $t$ that will be useful throughout the paper is 
\begin{equation}\label{ex:utility}
V_p(t-A_p+1) =  D_p(|S_p(t)|) - C_p(d_p(t) - A_p+1),\quad \forall t \ge A_p.
\end{equation}
where $D_p$ is a utility function that is naturally expected to follow law of diminishing returns, i.e., the incremental utility in transmitting more sub-packets decreases as the number of transmitted sub-packets increase. In particular we assume $D_p$ to be sub-modular (Definition \ref{defn:submod}). A popular example for $D$ that is sub-modular is $D(\ell) = 2^{(k_p-\ell)b}$ \cite{vaze2017energy,orhan2015source}. Importantly, sub-modular functions also include linear functions.
Function $C_p$ accounts for the usual delay metric, or the more modern age metric. Typically $C_p$ is assumed to be convex, e.g., $C_p(d_p(t) - A_p+1) = c_p (d_p(t) - A_p+1)$ a linear delay cost.

For each packet $p$, the 'effective' utility is $$V_p = \lim_{t\rightarrow \infty}V_p(t-A_p+1).$$
\begin{rem}
We can even allow individual hard packet deadlines in this model, by making $V_p(t-A_p) = 0$ if $d_p(t) > c_p$ if the deadline for packet $p$ is $c_p$. All the results presented in this paper apply when packets have individual hard deadlines, but we suppress the extra notation required to state this everywhere for the ease of exposition.
\end{rem}
In each slot, if $k$ sub-packets are transmitted simultaneously, the energy cost is assumed to be $g(k)$, where $g(.)$ is convex function with $g(0)=0$. For example, $g(k) = 2^{k\sfb}-1$ using the Shannon formula.

\begin{problem}\label{prob:obj}
For arbitrary positive weights $w_p$, the overall objective function is 
\begin{equation}
V =  \sum_{p \in \cP(t)} w_p V_p - \sum_{t}g(x_t),
\end{equation}
\end{problem}
\noindent where in slot $t$, $x_t$ sub-packets (possibly corresponding to different packets) are transmitted.
The problem is to maximize $V$ with respect to $x_t$ and which sub-packets to transmit among the packets that have arrived till $t$, for each time slot $t$. We call Problem \ref{prob:obj}, as age and quality of information (AQI), since it models the tradeoff between the two important metrics, age/delay and the distortion (quality) for packets/sources.

\begin{rem}\label{rem:indexing} For allowing packet splitting and assigning a distortion metric with respect to the number of transmitted sub-packets $D_p(|S_p|)$, we are essentially assuming that a compression algorithm is used to create a courser-to-finer quantization of packets into sub-packets, with the increasing index of sub-packets. 
Essentially sub-packet $i$ has utility $D(i)-D(i-1)$. Thus, sub-packet $1$ is most important and the importance level of sub-packets decreases with the sub-packet index, because of the sub-modular property.
This also ensures that the sub-packets are delivered in order, i.e., sub-packet $i$ is always delivered before sub-packet $j, j >i$ since the utility (importance) of $i$ is more than that of sub-packet $j$.

\end{rem}

We consider the online setting to solve Problem \ref{prob:obj}, where the algorithm has to make a causal decision about $x_t$ and which sub-packets to transmit at time $t$, and where the future packet arrivals can be adversarial. The performance metric for online algorithms is called the competitive ratio, that is defined for an algorithm $A$ as 
\begin{equation}\label{def:compratio}
\sfr_A = \min_{\sigma} \frac{V_A(\sigma)}{V_\opt(\sigma)}, 
\end{equation}
where $\sigma = \{\cP(t)\}$ is the input sequence of packets, and $\opt$ is the optimal offline algorithm (unknown) that knows the input sequence in advance. The objective is to find $A^*$ that maximizes $\sfr_A$. This setup appears too pessimistic, however, in the sequel, we show that online algorithms with competitive ratio of at most $2$ are possible.

\begin{definition}\label{defn:submod}
Let $N$ be a finite set, and let $2^N$ be the power set of $N$. A real-valued set function $f:2^N\to\mathbb{R}$ is said to be {\it monotone} if $f(S) \le f(T)$ for $S \subseteq T \subseteq N$,  and {\it sub-modular} if  $
f(S \cup \{i\}) -  f(S)\ge  f(T \cup \{i\}) - f(T)$
for every $S\subseteq T\subset N$ and every $ j \notin T$. 
\end{definition}

\section{Binary AQI problem}
In this section, we consider a special case of the AQI problem, called the binary AQI problem, where each packet $p$ has $b$ bits, i.e. $k_p=1$. Even in this setting, the AQI problem, is non-trivial, where the decision that the online algorithm has to make is to whether transmit a particular packet $p$ at all, and if yes, in which slot. 

In order to solve Binary AQI problem, we introduce a more general problem, called the online maximum-weight matching with vertex locking as follows.

\begin{problem}[Online Maximum-Weight Matching with Vertex Locking] \label{prob:OnlineMatchingWithLocking}
Given a bipartite graph $G = (A \cup B, E)$ where left nodes $a\in A$ arrive sequentially in time (arbitrary order) and right nodes $B$ are available ahead of time. On its arrival, $a$ reveals the weight $w_{ab}\ge 0$ for edges $\{(a,b) : b \in B\}$. Each node $b\in B$ has a locking time $T_b$, i.e., if $b$ is not matched to any $a\in A$ by then, it cannot be matched thereafter, and if some $a\in A$ is matched to $b$ at time $T_b$, then edge $(a,b)$ is always part of the (final) matching. Until time $T_b$ the left node matched to $b$ can be changed. Under these constraints, the objective is to construct an online matching from $A$ to $B$ having maximum weight, where $a$ can be matched to some $b$ any time after its arrival, and also the matched edge to $a$ can be changed to any of the unlocked $b$'s at any time. 

%
\end{problem}

Problem \ref{prob:obj} under the binary case $k_p=1 \ \forall \ p$ is an instance of Problem \ref{prob:OnlineMatchingWithLocking} by noting the following. \begin{enumerate}
\item The set of left nodes ($A$) are the set of packets $P$.
\item The set of right nodes are energy sub-slots $\{b_{t,i},\ i =1,2,\dots \}$ for each time slot $t$, where $i$ represents the (energy sub-slots) number of packets that are simultaneously transmitted in the same time slot. Matching any packet $p$ to vertex $b_{t,i}$ means that $p$ is transmitted in time-slot $t$ with $i-1$ other packets, thus incurring an additional incremental energy cost of $g(i)-g(i-1)$. The edge $e$ between packet $p \in P$ and $b_{t,i}$ has weight:
$w_e = V_p(t-T_p) - \left[g(i) - g(i-1) \right]$.
\item The set of vertices $\{b_{t,i},\ i =1,2,\dots \}$ lock at time $t$, corresponding to expiration of time slot $t$.	
\end{enumerate}
\begin{rem}We have modelled our binary AQI problem as an online bipartite matching problem, however, it is worth noting that no online maximum weight matching algorithm has a bounded competitive ratio when the edge weights are arbitrary \cite{secretary, korulapal, kesselheim}. To derive online algorithms with bounded competitive ratio \cite{korulapal, kesselheim} a secretarial model of input is assumed, where left side nodes arrive in a uniformly random order. We avoid the secretarial input model restriction by exploiting the exact utility function that governs the edge weights, and show that an online greedy algorithm is $2$-competitive.
\end{rem}
We propose Algorithm \ref{alg:matching-algorithm} to solve Problem \ref{prob:OnlineMatchingWithLocking} that maintains a temporary matching 
$\mathsf{TEMP}$ at every time instant that is the maximum-weight matching between unlocked (not matched to locked vertices of $B$) vertices of $A$ with the unlocked vertices in $B$. When any $b\in B$ locks, edge from $\mathsf{TEMP}$ that is matched to $b$ is added to the final output matching, $\mathsf{PERM}$. Every time a new vertex in $A$ arrives, $\mathsf{TEMP}$ is recomputed.

\begin{algorithm}[H]
\caption{}
\label{alg:matching-algorithm}
\begin{algorithmic}[1]
    \State \textbf{Initialize:} $\mathsf{PERM} =\phi$, $\mathsf{TEMP} = \phi$
    \State{All nodes of $B$ are unlocked}
    \While{$t \le T$}
    \State{On arrival of vertex $a \in A$ at time slot $t$ }
    \State{$a $ is defined to be unlocked}
    \State $\mathsf{TEMP} \gets$ maximum weight matching between unlocked nodes of 
    $A$ and unlocked nodes of $B$.
    \State{If vertex $b \in B$ locks at time slot $t$,}
    \If{$\exists a' : (a',b) \in \mathsf{TEMP}$}
    \State{$\mathsf{PERM} \gets \mathsf{PERM} \cup \{(a',b)\}$}
    \State{$a'$ is locked}
	\EndIf
    \EndWhile
    \State \textbf{Return} $\mathsf{PERM}$
\end{algorithmic}
\end{algorithm}

In the context of Problem \ref{prob:obj}, on arrival of each new packet, Algorithm \ref{alg:matching-algorithm} finds a temporary matching between the outstanding packets, and the current and future energy sub-slots, and makes the edges in the temporary matching permanent that are incident on the expiring time slot. We have the following result for Algorithm \ref{alg:matching-algorithm}.

\begin{theorem}\label{thm:binary} For input $\sigma$, let the offline-optimal matching to Problem \ref{prob:OnlineMatchingWithLocking} as $\opt$ with weight $W_{\opt}$ and the solution generated by Algorithm \ref{alg:matching-algorithm} as $\mathsf{PERM}$ with weight $W_{\mathsf{PERM}}$, we have 
\begin{equation*}
 W_{\mathsf{PERM}}(\sigma) \ge \frac{W_{\opt}(\sigma)}{2}.
\end{equation*}
\end{theorem}
The proof of Theorem \ref{thm:binary} is very different from usual online matching algorithms e.g. \cite{korulapal, kesselheim} mainly because of the fact that arriving packets can be sent in any time slot in future and scheduling decisions can also be deferred to future slots.

Thus, Algorithm \ref{alg:matching-algorithm} is $2$-competitive for solving the binary AQI problem with the worst case and does not require enforcing the secretarial input. 
Even though Algorithm \ref{alg:matching-algorithm} has an intuitive appeal it fails to provide a feasible solution for the general AQI problem as can be seen by the following example. Let a packet $p$ that arrives at time slot $0$ have two sub-packets $s_1$ and $s_2$. Following \eqref{ex:utility}, the delay for packet $p$ is delay of $s_1$ plus the incremental delay for $s_2$ starting from $s_1$ (if $s_2$ is transmitted) otherwise its delay of $s_1$ itself. 
Thus, if $s_1$ is matched by Algorithm \ref{alg:matching-algorithm} to node $b_{t,i}$ for some $i$, i.e., it is transmitted in slot $t$. 
Then to reflect the right delay cost 
(for example \eqref{ex:utility}) for packet $p$, the weight of edges from $s_2$ have to be changed (to account for incremental delay cost from $s_1$) depending on the choice made by  Algorithm \ref{alg:matching-algorithm}. Thus, dynamically, edge weights have to be changed depending on the earlier scheduling choices made by the Algorithm \ref{alg:matching-algorithm}, making it hard to compare it against $\opt$.
Thus, in the next section, we propose a more general online problem to model the general AQI problem and derive its competitive ratio.

\section{General AQI Problem}
In this section, to solve Problem \ref{prob:obj}, we introduce a more general problem (Online Resource Allocation with Locking (Problem \ref{RA:locking})) and provide a $2$-competitive algorithm for that. 

\begin{problem}[Online Resource Allocation Problem with Locking] \label{RA:locking}
Resources $r\in R$ arrive online at time $T_r$ and a set of bins $B$ are available before hand. The valuation of allocating $r$ to $b \in B$ is given by a monotone and submodular set-function 
$Z(S) : 2^{R \times B} \to \mathbb{R}^+$ is given, with increments $\rho(r,b|S) := Z(S \cup \{(r,b)\}) - Z(S)$.
The following properties of the problem make it very  different from the standard online sub-modular maximization (Problem \ref{prob:stand}) because : i) any resource $r \in R$ can be allocated to any bin at any time after its arrival and not necessarily on its arrival, ii) each bin $b \in B$ locks at an arbitrary time $T_b$. Once a bin $b$ locks, no new resources from $R$ can be allocated to or removed from $b$ from that time onwards, iii) the allocation of $r\in R$ to a bin is revocable, i.e., allowed to be changed to any other bin that has not been locked by then. Finally, any $r\in R$ can be discarded at any time, i.e., not allocated to any bin in which  case it accumulates a value of $0$. The objective is to find an online allocation of resources in $R$ to bins $B$ that maximizes $Z(S)$ under the above defined constraints. 

\end{problem}

We now show that Problem \ref{prob:obj} is a special case of Problem \ref{RA:locking}  as follows. \begin{enumerate}
\item The set of resources ($R$) are the set of all sub-packets $\cup_{p \in \cP(t)} \{p_i,\ i=1,2,\dots,k_p\}$. When packet $p$ arrives, $k_p$ resources arrive,  resource $r_{p_i}$ corresponds to sub-packet $p_i$.
\item There is a bin $b_t$ for each time slot $t$. Allocating multiple resources to $b_t$ indicates that the corresponding sub-packets are scheduled to be transmitted together at time slot $t$.
\item Let $S = \{(p_i,b_{t_i}),\ i=1,2,\dots\}$, where each $(p_i,b_{t_i})$ pair indicates that sub-packet $p_i$ of packet $p$ is transmitted at time $t_i$, and $Z(S)$ is defined as :
\begin{equation} \label{eq:Z(S)-val}
Z(S) = \sum_{p \in P} D_p(|S_p|) - \sum_{p \in P}C_p(d_p - A_p) - \sum_t g(|S_{b_t}|)
\end{equation}
to match \eqref{ex:utility}, where $|S_{b_t}|$ is the number of sub-packets allocated to bin $b_t$ (time slot $t$) in $S$, $|S_p|$ is the number of sub-packets of packet $p$ that have been allocated in $S$ (that will be transmitted) and $d_p$ is the earliest time slot after which no sub-packet of packet $p$ is allocated in $S$. 
It easily follows that $Z(S)$ is monotone and sub-modular, since $D_p$ is a monotone and sub-modular function, and $C_p$ and $g$ are increasing and convex functions. The incremental valuation of $Z(S)$ also follows from (\ref{eq:Z(S)-val}) explicitly. In the solution $S$, let $d_p^S$ denote the time at which the last subpacket of packet $p$ is scheduled. Then, the incremental valuation of $Z(S)$ when subpacket $p_j$ of packet $p$ is allocated at time $t_j$ is:
\begin{equation} 
\begin{split} \rho (p_j,b_{t_j} | S) =
&\Delta D_p (|S_p|) - \Delta g(|S_{b_{t_j}}|) \\  &- [C_p (\max \{ d_p^S, t_j \} - A_p) \\\label{eq:incremental-val} & - C_p (d_p^S - A_p)]
\end{split}
\end{equation}
where $\Delta f (x) = f (x + 1) - f(x)$.
\item At the end of each time-slot $t$, the bin $b_t$ locks. This reflects the causality constraint that once the time-slot $t$ expires, sub-packets already transmitted in that slot cannot be undone or transmitted again, and no new sub-packets can be transmitted in time slot $t$.
\end{enumerate}
\begin{rem} As discussed earlier, the binary AQI problem cannot handle the case when there is more than one sub-packet for any packet, since it is limited in its ability to count incremental delays for sub-packets.
Problem \ref{RA:locking} allows the flexibility of the delay associated with serving a sub-packet at some time $t$ to depend on the last time a sub-packet from the same packet was served. To illustrate this point, for instance, consider the case when some packet $p$ having two sub-packets $s_1$ and $s_2$, and WLOG assume that $s_1$ is scheduled at time $t_1$, where the running solution of packet-time slot allocation is $S$. Then the increment in valuation of $S$ upon processing subpacket $s_1$ at time $t_1$, from (\ref{eq:incremental-val}), would be equal to $ \rho(s_1,b_{t_1}|S) = D_p (1) - \Delta g (|S_{b_{t_1}}|) - C_p (t_1 - A_p)$. Here, the total delay of packet $p$ is accounted for, since $\rho (s_1, b_{t_1} | S)$ contains the term $C_p(t_1-A_p)$. Scheduling $s_1$ at time $t_1$ augments the current solution $S$ with the tuple $(s_1, b_{t_1})$, and the solution at any future point $S'$, is guaranteed to include both $S$ and $(s_1,b_{t_1})$. If subpacket $s_2$ is scheduled at a later time $t_2$, the increment in valuation of $S'$, again follows from (\ref{eq:incremental-val}): $\rho(s_2,b_{t_2}|S') = D_p (2) - D_p (1) - \Delta g(|S_{b_{t_2}}|) - [C_p (t_2 - A_p) - C_p (t_1 - A_p)]$. The term $C_p (t_2 - A_p) - C_p (t_1 - A_p)$ captures the incremental delay of $s_2$ starting from $s_1$ which is implicitly dependent on $t_1$, the time at which a subpacket from $p$ is transmitted. Thus the overall delay faced by packet $p$ is captured in two parts: the delay of $s_1$, and the incremental delay of $s_2$ starting from $s_1$. 
The flexibility of $Z(S)$ in Problem \ref{RA:locking} allows us to define $\rho(s_2,b_{t_2}|S')$ this way, to include $C_p(t_2-A_p) - C_p(t_1-A_p)$ which 
is implicitly dependent on $t_1$. 
\end{rem}


We propose the following greedy algorithm (Algorithm \ref{alg:on-greedy}) to solve Problem \ref{RA:locking}, where $S \subseteq R \times B$, and $\rho(r,b|S) := Z(S \cup \{(r,b)\}) - Z(S)$. 
Note that Algorithm \ref{alg:on-greedy} makes irrevocable resource-bin allocations even though Problem \ref{RA:locking} allows revocable allocations to unlocked bins.
\begin{algorithm}[H]
\caption{Greedy Algorithm for Problem \ref{RA:locking}} \label{alg:on-greedy}
\begin{algorithmic}[1]
\Procedure{$\mathsf{GREEDY}$}{}
    \State \textbf{Initialize:} $G^0=\phi$, $i=0$
    \While{$t \le T$}
    \If{at some time $t$, resource $r$ arrives}
    \State{Irrevocably allocate $r$ to bin $b$ }
    \State{if $b^\star = \underset{b \in B_t}{\text{argmax}} \left\{\rho(r,b|G^{i-1}) \right\}$}\\
    \Comment{Tie: broken arbitrarily}
    \State $G^i \gets G^{i-1} \cup \{(r,b^\star)\}$, $i \gets i+1$
    \EndIf
    \EndWhile
    \State \textbf{Return} $G=G^N$
\EndProcedure
\end{algorithmic}
\end{algorithm}
\begin{theorem} \label{theorem:12apx}
For Problem \ref{RA:locking}, with input $\sigma$, let the  online solution generated by Algorithm \ref{alg:on-greedy} be $G$ and the optimal-offline solution be $\opt$, then we have that:
\begin{equation*}
Z(G)\ge \frac{ Z(\opt)}{2}.
\end{equation*}
\end{theorem}
Thus, Theorem \ref{theorem:12apx} shows that the greedy algorithm similar to the standard sub-modular maximization \cite{Fisher1978} achieves a competitive ratio of at least $1/2$ for Problem \ref{RA:locking} even though by definition it has more attributes and more flexibility than the standard sub-modular maximization. In the context of AQI problem, which as shown before is a special case of Problem \ref{RA:locking}, this is significant since AQI problem has many different combinatorial elements, and the cost functions $D_p, C_p$ and $g$ are fairly generic. In the next section (Section \ref{sec:specialcases}), we show that the specific instances of AQI problem can be used to model various well studied problems in literature, and guarantees derived in Theorem \ref{theorem:12apx} carry over to them, which were previously not known.

\section{Special Cases of the AQI Problem}\label{sec:specialcases}
In this section, we describe that three important problems, as discussed in the Introduction, can be closely modelled by the AQI problem.
\subsection{Age of Information with Multiple Sources}\label{subsec:aoi}

An AoI problem with $\sfs$ sources was considered in \cite{kadotascheduling}, that generalizes the single user AoI metric, where there are multiple sources, and at any time information about only one of the sources can be transmitted to the monitor. For a fixed source, if event/update $i$ happens at time $t_i$ which is made available at the receiver at time $t$, then age for update $i$ is $\Delta_i = t-t_i$, and overall age is the time average of $\Delta_i$'s, area under the curve as shown in Fig. \ref{fig:sawtooth} normalized by time horizon. 

\begin{figure}[H]
	\centering
	\begin{tikzpicture}[bipartite,every text node part/.style={align=left}]
	\node[xshift=6cm,yshift=-4cm]  {$\Delta_1 = t-t_2$ $\Delta_2 = t+2-t_3$};
	\node[xshift=4cm,yshift=0.6cm]  {Timeline of events from $s_1$};
		\begin{scope}[xshift=2.25cm,start chain=going right,node distance=3mm]
		\foreach \i in {3,...,1}
			\node[on chain] (s1\i) {$t_\i$};
		\end{scope}
		\node[xshift=8cm,yshift=0.65cm,label=right:$\quad$Age] (oldage) {};
		\node[xshift=8cm,label=right:$\quad \Delta_1$] (t0) {$t$};
		\node[xshift=8cm,yshift=-0.75cm,label=right:$\Delta_1+1$] (t1) {$t+1$};
		\node[xshift=8cm,yshift=-1.5cm,label=right:$ \Delta_2$] (t2) {$t+2$};
		\begin{scope}[xshift=8cm,yshift=-2.25cm,start chain=going below,node distance=0.25cm]
		\foreach \i in {3,4}
			\pgfmathsetmacro\delay{\i-2}
			\node[on chain,label=right:$\Delta_2+$\pgfmathprintnumber{\delay}] (t\i) {$t+\i$};
		\end{scope}
		\draw (s11) -- (t0); 
		\path[draw,dotted] (s13) edge[out=330,in=180] node[centerlabel,pos=0.65]{$0$}  (t1);
		\path[draw] (s12) edge[out=280,in=170] node[centerlabel,pos=0.6]{\footnotesize $D_{s_1}-(2\Delta_1+2)$}  (t2);
		\path[draw,dotted] (s13) edge[out=270,in=185] node[centerlabel,pos=0.65]{$0$}  (t2);
		\path[draw,dotted] (s13) edge[out=270,in=180] node[centerlabel,pos=0.75]{\footnotesize $D_{s_1}-(\Delta_2+0.5)$}  (t3);
		\path[draw,dotted] (s13) edge[out=270,in=180] node[centerlabel,pos=0.75]{\footnotesize $D_{s_1}-(2\Delta_2+2)$}  (t4);
	\end{tikzpicture}
	\caption{$\rho(r,b|S)$ for source $s_1$ and $b=b_t,b_{t+1},\dots$ (Solid edges $\in S$, dotted edges $\not\in S$)}
	\label{fig:new}
\end{figure}
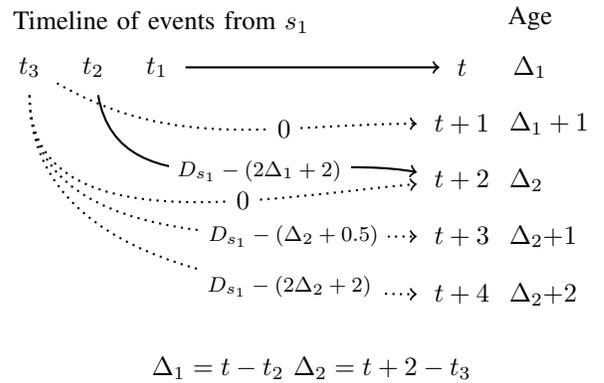

Consider a source $s_1$ that generates events at times $\{t_i, i=1,2,\dots\}$.  
We use $e_i$ to represent the event occurring at time $t_i$, for example $i=1,2,3$ as shown in Fig. \ref{fig:new}. Let $e_1$ and $e_2$ be scheduled to be transmitted at slot $t$ and $t+2$ respectively,  as shown by solid lines in Fig. \ref{fig:new}. Given this allocation, we next define the edge weights ($\rho(e_{t_i},b_t|S)$) in Fig. \ref{fig:new} between event $e_3$ and multiple slots as dotted lines to correctly reflect the age metric. 
Recall that for the corresponding AQI problem, the objective is to maximize the difference between the total value (accrued by processing events) and the average age of information. 
For simplicity, let the value accrued by transmitting any of the events $e_i$ for source $s_1$ be $D_{s_1}$.

Since $e_2$ is already scheduled to be transmitted at slot $t+2$, scheduling $e_3$ to be processed at slot before or at $t+2$ does not increase the age. Moreover, $\rho(e_3,b_{k}|S)  = 0$ for all $k\le t+2$, since if $e_3$ is sent before $e_2$, $e_2$ will be outdated and accrue no value. To be precise, $e_2$ accrues no value if $e_3$ is sent before $e_2$, so without loss of generality, we let the increment  $\rho(e_3,b_{k}|S)  = 0$ for all $k\le t+2$ over the previous allocation $S$. Thus the edge weights from $e_3$ to slot $ t+1$, and $t+2$ are zero. Scheduling $e_3$ at slot $t+3$ or $t+4$ increases the area under the sawtooth curve (Fig. \ref{fig:sawtooth}) by $\Delta_2+0.5$ (blue area) and $2\Delta_2+2$ (blue + green area) respectively. Therefore, $\rho(e_3,b_{t+3}|S)$ is defined as $D_{s_1} - \Delta_2 + 0.5$ and $\rho(e_3,b_{t+3}|S)$ is defined as $D_{s_1} - (2\Delta_2 + 2)$.

\begin{figure}[H]
\centering
\begin{tikzpicture}[every text node part/.style={align=left}, scale=0.75]
\draw[->] (-1,0) -- (6,0); 
\draw[->] (-1,-0.5) -- (6,-0.5);
\draw (1,-0.5) node[anchor=north] {\footnotesize $t$}
		 (2,-0.5) node[anchor=north] {\footnotesize $t+1$}
		 (3,-0.5) node[anchor=north] {\footnotesize $t+2$}
		 (4,-0.5) node[anchor=north] {\footnotesize $t+3$}
		 (5,-0.5) node[anchor=north] {\footnotesize $t+4$};

\draw[dotted] (5,0.5) -- (0,0.5) node[anchor=east] {$\Delta_1$};
\draw[dotted] (5,1.25) -- (0,1.25) node[anchor=east] {$\Delta_2$};
\draw[dotted] (-0.75,0) node[anchor=north] {$t_1$};

\draw[->] (0,-0.5) -- (0,4); 
\draw[thick] (0,0.75) -- (1,1.75) -- (1,0.5) -- (3,2.5) -- (3,1.25) -- (5,3.25);
\draw[dotted] (1,0.5) -- (0.5,0) node[anchor=north] {$t_2$};
\draw[dotted] (1,1) -- (1,-0.5); 
\draw[dotted] (2,1.5) -- (2,-0.5);
\draw[dotted] (3,1.25) -- (1.75,0) node[anchor=north] {$t_3$};
\draw[dotted] (3,1.75) -- (3,-0.5);
\draw[dotted] (4,2.25) -- (4,-0.5);
\draw[dotted] (5,3.25) -- (5,-0.5);
\draw[dotted] (3,1.75) -- (3,-0.5);
\draw[dotted] (0,0.75) -- (-0.75,0);
\draw[dotted,fill=blue,fill opacity=0.2] (3,0) -- (3,1.25) -- (4,2.25) -- (4,0);
\draw[blue!80!black] (3.5,1) -- (3.5,2.75) node[anchor=south] {Area $=$ \\ \footnotesize $\Delta_2+0.5$};
\draw[green!50!black] (4.5,1.5) -- (5.5,1.5) node[anchor=west] {Area $=$ \\ \footnotesize $\Delta_2+1.5$};
\draw[dotted,fill=green,fill opacity=0.2] (3,0) -- (3,1.25) -- (5,3.25) -- (5,0);
\end{tikzpicture}
\caption{Age of Information ($\Delta_2 = t+2-t_3$)}
\label{fig:sawtooth}
\end{figure}
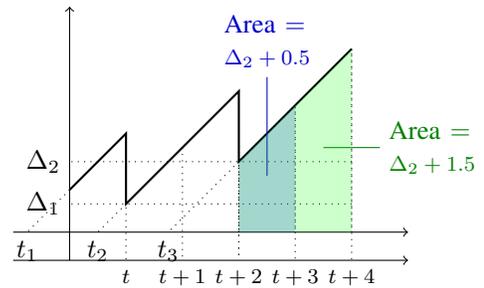

Thus the AQI problem is a generalization of the AoI problem \cite{kadotascheduling}, where the general delay cost $C(.)$ can model the age metric for each of the sources, while the throughput constraint is inherently captured via the distortion function. The additional feature of the AQI problem is the inclusion of the energy cost, which captures the physical limits on the speed of transmission.

\subsection{Remote Sampling}
A remote sampling problem is considered in \cite{DBLP:journals/corr/SunPU17}, where an observer measures a single  Wiener process that it wants to be reproduced at the receiver/monitor under a sampling frequency constraint. The channel between the observer and monitor is a random delay channel, and first-in-first-out schedule is enforced for the transmission queue. Let $W_t$ be the true value of the Wiener process at time $t$, and ${\hat W}_t$ be the reproduced value at the monitor, the objective is to minimize the MMSE $\frac{1}{T}\bbE\left\{\int_0^T (W_t - {\hat W}_t)^2\right\}$ subject to a sampling constraint at the observer.

AQI problem takes a slotted view of this remote estimation problem, where the packet distortion $(W_t - {\hat W}_t)^2$ is captured by how many sub-packets of each packets are transmitted (function $D_p$), and the delay is accounted explicitly by counting the difference between the arrival of the packet and the reception of its last transmitted sub-packet. Moreover, the sampling frequency constraint is enforced implicitly since there are packets arriving from multiple sources and energy cost is convex. Thus each sources' packets get a transmission chance only with a fixed maximum rate. AQI problem, is however, more general, since packets can be generated from multiple sources, and more importantly, no constraints are enforced on source distribution. The source samples can be generated arbitrarily or even by an adversary. In comparison to \cite{DBLP:journals/corr/SunPU17} which derives an optimal algorithm for the Wiener process, the $\mathsf{GREEDY}$ algorithm for the AQI problem is $2$-competitive for this general setting.

\subsection{Speed Scaling with Multiple Servers} 
We show in this subsection that AQI problem can model the  
classical speed scaling problem \cite{yao1995scheduling, bansal2009speedconf, wierman2009power, gupta2010scalably, gupta2012scheduling} with multiple servers.
With speed scaling, jobs arrive over time and on its arrival each job has to be assigned to one of the multiple servers, where the server speed is variable, and running at speed $s$ incurs a cost of $P_i(s)$ for server $i$. Both preemption and job-migration is also allowed, i.e., a single job can be processed by different servers at different times.

The AQI problem can model a slotted version of the speed scaling problem with multiple servers, where the speed once chosen is fixed for a time slot, while in the usual speed scaling problem speed can be changed continuously. 
For modelling the speed scaling problem with multiple servers $m$ by the AQI problem, we only need to replace the energy function $g(k)$ by $g_i(k)$ (convex function) for each of the $i=1,\dots,m$ servers that plays the role of $P_i(s)$. One important point to note is that, with the AQI problem, multiple sub-packets belonging to the same packet can be served by multiple servers at the same time, which is typically not allowed with the classical speed scaling problem. However, the best known algorithms for speed scaling with multiple servers \cite{gupta2010scalably, gupta2012scheduling}, use  processor sharing, that also 'inherently' violate this constraint, since with processor sharing using multiple servers, a fixed number of jobs are processed by all the processors in each small time interval. 

It is easy to follow that with $m$ servers, each having energy function $g_i(k)$, Problem \ref{RA:locking} just has $m$ times more bins, and algorithm $\mathsf{GREEDY}$ remains unchanged and its guarantee derived in Theorem \ref{theorem:12apx} still holds.
Thus, the $\mathsf{GREEDY}$ algorithm for the AQI problem that uses job splitting similar to processor sharing of \cite{gupta2010scalably, gupta2012scheduling} is a $2$-competitive algorithm for the speed scaling problem with multiple servers, which is far better than the best known results of  \cite{gupta2010scalably, gupta2012scheduling}, that are either $P(.)$ dependent or need speed augmentation compared the $\opt$. An important point to note is that with classical speed scaling all packets have to be transmitted. AQI problem allows the flexibility of dropping certain packets if they are not profitable. The constraint that all packets have to be transmitted can also be enforced with AQI problem by controlling weights $w_p$.

\section{Conclusions}
In this paper, we have introduced a resource allocation problem (called AQI problem) that captures the tradeoff between three important cost metrics, delay/age, distortion, and energy. The actual cost function is a linear combination of the three costs, and under a worst-case model, an online algorithm is proposed that is $2$-competitive. The AQI problem is shown to be 'close' generalization of three important resource allocation problems, 
and the results derived in this paper provide new/improved results for the three problems.

\bibliographystyle{IEEEtran}
\bibliography{refs}
\appendices
\section{Proof of Theorem \ref{thm:binary}}
We follow the following notation for proving Theorem \ref{thm:binary}.
$A_t$ is defined as the nodes in $A$ that have arrived by time $t$.
$T_b$ is the time at which vertex $b \in B$ is locked.
We denote the set of locked vertices in $B$ at time $t$ as $B_t^\perp$ and the set of vertices in $A$ matched to vertices in $B_t^\perp$ as $A_t^\perp$. These will be referred to as locked vertices in $A_t$. $A_t \setminus A_t^\perp$ are referred to as the unlocked vertices in $A_t$.
$L_t$ is defined as the set of edges that have been locked in (by virtue of their endpoints in $B$ having been locked) by time $t$.
$M(A_t, B, L_t)$ is the maximum weight matching from $A_t$ to $B$ with the condition that the matching must include the edges in $L_t$. Referring to Algorithm $\ref{alg:matching-algorithm}$, it follows that at any time, $\mathsf{TEMP}$, by definition is equal to $M(A_t, B, L_t) \setminus L_t$.
$\nu_b$ is defined as the weight of the edge matched to $b \in B$ in the final matching generated by the algorithm.
For any vertex $b \in B$ and time $t$, define $\rho_t(b)$ as follows:
        \begin{equation} \label{eq:rho-def}
    \rho_t(b) = \begin{cases}
    W(A_t,B,L_t) - W(A_t,B \setminus \{b\},L_t) \quad & t < T_b, \\
    \nu_b & t \ge T_b.
    \end{cases}
    \end{equation}
If $b$ is unlocked at time $t$, $\rho_t(b)$ is defined as $W(A_t, B, L_t) - W(A_t, B\setminus \{b\}, L_t)$. It is the difference in weights of the maximum matching of $A_t$ to $B$ and $A_t$ to $B \setminus \{b\}$ conditioned on $L_t$ being a subset of both matchings.
If $b$ is locked at time $t$, then $\rho_t(b)$ is defined as $\nu_b$, the weight of the edge incident on $b$ in $L_t$.
Suppose that node $a \in A$ arrives at time $t$. Define $\Delta_a := W(A_t,B,L_t) - W(A_{t-1},B,L_{t-1})$, the change in the weight of $\mathsf{TEMP}$ due to this arrival.
$T$ is the time horizon, i.e. the time at which all vertices in $B$ are assumed to be locked. We assume that no vertex in $A$ arrives after time $t$.

\begin{proof}
At any time $t$, assume that some $a \in A$ arrives, and consider arbitrary bin $b \in B$ that is not yet locked (at least one unlocked bin exists from the definition of $T$). Recall that if at time $t$ resource $a$ arrives, $\Delta_a = W(A_t,B,L_t) - W(A_{t-1},B,L_{t-1})$, and $\rho_t(b) = W(A_t,B,L_t) - W(A_t,B \setminus \{b\},L_t)$ (since $b$ has not locked yet). Therefore, we have: $\Delta_a + \rho_{t-1}(b)$
    \begin{align}
         &= \left[W(A_t,B,L_t) - W(A_{t-1},B,L_t)\right] + \\ \nn
         &  \ \ \ \ \ \left[W(A_{t-1},B,L_t) - W(A_{t-1},B \setminus \{b\},L_t) \right], \nonumber\\
        &= W(A_t,B,L_t) - W(A_{t-1}, B \setminus \{b\}, L_t). \label{eq:005}
    \end{align}
    The RHS of (\ref{eq:005}) must be at least as much as the weight of the edge between $a$ and $b$, $w_{ab}$, since a way to form the matching $M(A_t,B,L_t)$ is to form the (partial) matching $M(A_{t-1},B \setminus \{b\},L_t)$ and augment the edge $(a,b)$. Therefore, 
    \begin{align}
        &\Delta_a + \rho_{t-1}(b) \ge w_{ab}, \nonumber \\
        \Rightarrow\ &\Delta_a + \nu_b \overset{(i)}{\ge} w_{ab}, \label{eq:006}
    \end{align}
    where $(i)$ follows from Lemma \ref{lemma:rho-inc}: $\rho_{t-1}(b) \le \rho_{T}(b) = \nu_b$. Inequality (\ref{eq:006}) holds for any arbitrary $b \in B$ that is unlocked at time $t$, in particular for the vertex $b^*$ to which $a$ is matched in the offline-optimal solution. Summing (\ref{eq:006}) over all $(a,b^*)$ pairs in the offline-optimal solution, and noting that $\sum_{a \in A} \Delta_a = \sum_{b \in B} \nu_b = W_{ALG}$,
    \begin{equation*}
    2W_{ALG} \ge W_{OPT}.
    \end{equation*}
\end{proof}

\begin{lemma} \label{lemma:rho-inc}
    $\rho_t(b)$ defined in (\ref{eq:rho-def}) is a non-decreasing function of $t$ for any vertex $b$ in $B$:
    $$ \forall t \in [0,T-1],\ \rho_t(b) \le \rho_{t+1}(b).$$
    Intuitively, this means that as time progresses, the utility of an unlocked vertex $b$ increases until the time it is locked, and once it is locked by definition anyway it has fixed utility $\nu_b$.
\end{lemma}
The intuition for Lemma \ref{lemma:rho-inc} is that if vertex $b$ is added  
to the set of right vertices, then at each time $t$ till $b$ locks, a dynamic 
temporary matching matches some vertex $a_t$ to $b$. If all packets arrive at time $0$, 
then $a_0$ matched to $b$ eventually becomes permanent, and there is no change to $\rho_t(b)$ with $t$. However, as more packets arrive over time (that is what essentially time progressing means in Lemma \ref{lemma:rho-inc} statement), the availability of $b$ allows the matching algorithm more flexibility in finding a matching with the largest weight, compared to when $b$ is not present. 
The rigorous proof of Lemma \ref{lemma:rho-inc} is provided in Appendix \ref{sec:Lem1}.

\section{Proof of Lemma \ref{lemma:rho-inc}}\label{sec:Lem1}
\begin{definition}[Flow network] \label{def:flow-network}
A flow network is a directed graph $G = (V,E)$ with distinguished vertices denoted $s$ (source) and $t$ (sink). Every edge $e \in E$ is associated with a weight $w_e$ and a capacity $c_e$, and is referred to as a \textit{$w_e$-weight $c_e$-capacity edge}.
\end{definition}

\begin{definition}[flow] \label{def:flow}
Given a flow network $G=(V,E)$, with weights $w_e$ and capacity $c_e$ on every edge $e \in E$, a flow $f = \{f_e : e \in E\}$ is a set satisfying the following properties:
\begin{enumerate}
\item $\forall e \in E,\ 0 \le f_e \le c_e$. The flow on an edge is positive and cannot exceed the capacity of that edge. If the flow through edge $e$ is equal to $c_e$, it is said to be saturated.
\item For an edge $e = (u,v) \in E$, we use $t(e)$ to denote $u$ and $h(e)$ to denote $v$. Then, 
\begin{equation*}
\forall v \in V \setminus \{s,t\},\ \sum_{e \in E : t(e) = v} f_e = \sum_{e \in E : h(e) = v} f_e
\end{equation*}
Given a flow $f$, $f_e$ is referred to as the flow through edge $e$. The quantity $\sum_{e \in E} f_e.w_e$ is referred to as the weight of the flow.
\end{enumerate}
\end{definition}

\begin{proof}[Proof of Lemma \ref{lemma:rho-inc}]
    For any vertex $b$, in case $t > T_b$, the proof is trivial, since $\rho_t(b)$ is defined as $\nu_b$ for any locked vertex and remains fixed throughout.
    For $t \le T_b$, the proof follows separately for 3 cases, where either:
    \paragraph{Vertex $b' \ne b$ is locked ($t < T_b$)} Since there is no arrival, $A_t = A_{t-1}$. The set of locked edges at time $t$, $L_t = L_{t-1} \cup e'$, where $e'$ is the edge in $L_{t}$ incident on $b'$ in the matching $M(A_{t-1},B,F_{t-1})$.
    By definition, since edge $e'$ is incident on $b'$ in the matching $M(A_{t-1},B,L_{t-1})$, it is the edge that is locked in at time $t$. The two matchings $M(A_{t-1},B,L_t)$ and $M(A_{t-1},B,L_{t-1})$ are identical. The only difference is that the edge $e'$ is a part of $L_t$ in $M(A_{t-1},B,L_t)$ since $b'$ locks at time $t$, but in $M(A_{t-1},B,L_{t-1})$ $e'$ is not included among the edges in $L_{t-1}$ since $b'$ has not locked yet. Therefore,
    \begin{equation} \label{eq:001} W(A_{t-1},B,L_t) = W(A_{t-1},B,L_{t-1})
    \end{equation}
    The matching $M(A_{t-1},B \setminus \{b\}, L_{t-1})$ is the optimal matching between $A_{t-1}$ and $B \setminus \{b\}$ when constrained to include $L_{t-1}$. Therefore, it cannot have lower weight than any matching between $A_{t-1}$ and $B \setminus \{b\}$ when constrained to include $L_t = L_{t-1} \cup \{e'\}$ - one such matching being $M(A_{t-1},B \setminus \{b\}, L_{t})$.
    \begin{equation} \label{eq:002}
        W(A_{t-1},B \setminus \{b\},L_t) \le W(A_{t-1},B \setminus \{b\}, L_{t-1})
    \end{equation}

Subtracting (\ref{eq:002}) from (\ref{eq:001}) and using the fact that $A_t = A_{t-1}$, from the definition of $\rho_t(b)$ the Lemma follows for this case.
    \paragraph{Vertex $b$ is locked ($t = T_b$)} Since $t=T_b$, the weight of the edge $e=(a',b)$ to $b$ in $M(A_{t-1},B, L_{t-1})$, $w_{a'b}$ is equal to $\nu_b$ (as this edge is locked at time $T_b$). A valid matching from the vertices in $A_{t-1}$ to $B \setminus \{b\}$ that includes $L_{t-1}$ is: $ M(A_{t-1}, B \setminus \{b\}, L_{t-1})$ with the edge $e$ removed and with $a'$ discarded. This matching has weight $W(A_{t-1},B,L_{t-1}) - w_{a'b} = W(A_{t-1},B,L_{t-1}) - \nu_b$. Since $M(A_{t-1}, B \setminus \{b\}, L_{t-1})$ is the optimal matching from $A_{t-1}$ to $B\setminus \{b\}$ that includes $L_{t-1}$, we have that:
    \begin{align*}
        &W(A_{t-1}, B \setminus \{b\}, L_{t-1}) \ge W(A_{t-1},B,L_{t-1}) - \nu_b \\
        \Rightarrow \ &\nu_b \ge W(A_{t-1},B,L_{t-1}) - W(A_{t-1}, B \setminus \{b\}, L_{t-1}) \\
        & = \rho_{t-1}(b), \\
        \Rightarrow \ &\rho_t(b) \ge \rho_{t-1}(b)
    \end{align*}
    \paragraph{Vertex $a \in A$ arrives} We begin with a brief proof outline. In order to prove the Lemma statement in this case, we construct a flow network $N_{12}$ and consider a flow $f_{12}$ on it with weight $W(A_t,B \setminus \{b\},L_t) + W(A_{t-1},B,L_{t-1})$. We then consider $f^*$, the maximum-weight flow over $N_{12}$ and show that it can be decomposed into valid flows, $F_1$ and $F_2$ such that the weight of $F_1$ is upper bounded by $W(A_t,B,L_t)$ and the weight of $F_2$ by $W(A_{t-1},B \setminus \{b\},L_{t-1})$. Using $H(f)$ to denote the weight of a flow $f$ over the network $N_{12}$, we have:
    $
W(A_{t},B \setminus \{b\},L_t) + W(A_{t-1},B,L_{t-1}) = H(f_{12})$
\begin{align*}
&\overset{(i)}{\le} H(f^*), \\
&= H(F_1) +H(F_2), \\
&\le W(A_{t},B,L_t) + W(A_{t-1},B \setminus \{b\},L_{t-1}),
\end{align*}
where $(i)$ follows by optimality of $f^*$.
This implies that $
W(A_{t-1},B,L_{t-1}) - W(A_{t-1},B \setminus \{b\},L_{t-1})$
\begin{align*} &\le W(A_{t},B,L_t) - W(A_{t},B \setminus \{b\},L_t).
\end{align*}
Therefore, 
$$ \rho_{t-1}(b) \le \rho_t(b).$$

In order to generate $N_{12}$, we first construct two flow networks $N_1$ and $N_2$. $N_{12}$ is then generated by appropriately superimposing $N_1$ and $N_2$. Since no vertices (and hence no edges) lock at time $t$, $L_t = L_{t-1}$. We also have that $A_t = A_{t-1} \cup \{a\}$. As shown in Figure \ref{fig:1c}, we build a flow network $N_1$ on the bipartite graph with vertices $A_t \cup B \setminus \{b\}$ and will consider a flow $f_1^*$ on it having weight $W(A_t,B \setminus \{b\},L_t)$ on it. $N_1$ is generated by first adding source $s_1$ and sink $t_1$, where $s_1$ is adjacent to every vertex in $A_t$ via $0$-weight $1$-capacity edges and $t_1$ is adjacent to every vertex in $B \setminus \{b\}$ through $0$-weight $1$-capacity edges. $B \setminus \{b\}$ is also augmented with $|A_t|$ dummy vertices, one for each vertex in $A_t$. Each vertex $a \in A_t$, is connected to its corresponding dummy vertex $d_a$ through a $0$-weight $1$-capacity edge. Each dummy vertex is also connected to $t_1$ through a $0$-weight $1$-capacity edge. $N_1$ now has $|A_t|$ left nodes and $|B|-1+ |A_t|$ right nodes. The flow $f_1^*$ is defined as the max-weight flow over $N_1$ under the constraint that every edge in $L_{t-1}$ must be saturated with unit flow. The presence of dummy nodes ensures that every vertex in $A_t$ is saturated with $1$ unit of flow, without affecting the value of the flow, which is $W(A_t,B \setminus \{b\},L_{t-1})$.
    
A similar procedure is followed to generate a flow network $N_2$ on the bipartite graph with vertices $A_{t-1} \cup B$. $N_2$ has $|A_{t-1}|$ left nodes and $|B|+|A_{t-1}|$ right nodes. We consider the max-weight flow over $N_2$ under the constraint that every edge in $L_{t-1}$ is saturated with unit flow and denote this as $f_2^*$. A similar argument as in the case of $f_1^*$ concludes that $f_2^*$ has weight equal to $W(A_{t-1},B,L_{t-1})$.

Flow-networks $N_1$ and $N_2$ are defined over a common set of vertices with the exception of $b$ (not in in $N_1$),  $a$ and $d_a$ (not in $N_2$) where $d_a$ is the dummy vertex corresponding to $a$. $N_1$ and $N_2$ are then superimposed to generate the network $N_{12}$ - each vertex is merged with its counterpart, and for edges between two such vertices, the capacity is doubled, but the weight is kept unchanged. The weights and capacities of edges that are incident on either $a, d_a$ or $b$ are kept unchanged. As shown in Figure \ref{fig:2a}, the vertex set of $N_{12}$ is $A_t \cup B \cup D \cup \{s_1,t_1\}$, where $D$ is the set of dummy nodes for $A_t$.

    Since $f_1^*$ is a valid flow on $N_1$ and $f_2^*$ is a valid flow on $N_2$, the superposition of $f_1^*$ and $f_2^*$ denoted $f_{12}$ having weight $W(A_t,B\setminus\{b\},L_t) + W(A_{t-1},B,L_t)$ is a valid flow on $N_{12}$. Furthermore, $f_{12}$ is constrained to have exactly $2$ units of flow through every edge in $L_t$ (since in both $f_1^*$ and $f_2^*$, every edge in $L_t$ has unit flow). Defining $f^*$ as the max-weight flow over $N_{12}$ with the constraint that every edge in $L_t$ has a flow of $2$, and recalling that $H(f)$ denotes the weight of flow $f$ over the network $N_{12}$, it follows that:
    \begin{align}
    H(f^*) &\ge W(A_t,B \setminus \{b\},L_t) + W(A_{t-1},B,L_t), \nonumber \\
    &\overset{(i)}{=} W(A_t,B \setminus \{b\},L_t) + W(A_{t-1},B,L_{t-1}), \label{eq:003}
    \end{align}
    where $(i)$ follows from the fact that no node is locked in at time $t$ and hence $L_t = L_{t-1}$. Due to the presence of dummy nodes having $0$-weight edges, there are multiple candidates for $f^*$ having the same total weight. One such candidate is the maximum weight flow on $N_{12}$ excluding the dummy vertices (every vertex except $a$ has a flow of $0$ or $2$, and $a$ has a flow of $0$ or $1$ through it). Henceforth, the maximum-weight flow described below is referred to as $f^*$: \begin{enumerate}
        \item the flow of every unsaturated (having flow $0$) vertex in $A_t \setminus \{a\}$ is saturated by passing $2$ units of flow through the $0$-weight edge to its dummy node.
        \item If $a \in A_t$ is unsaturated, unit flow is pushed through the $0$-weight $1$-capacity edge to its dummy node $d_a$.
    \end{enumerate}
    Every vertex in $A_t$ has $2$ units of flow through it in $f^*$, except for $a$, which has unit flow through it.
    
    \begin{figure*}[p]
	\begin{subfigure}[t]{0.29\linewidth}
	\centering
	\begin{tikzpicture}[bipartite]
		\begin{scope}[start chain=going below,node distance=3mm]
		\foreach \i in {1,2,...,4}
			\node[leftnode,on chain] (f\i) [label=left: $a_\i$] {};
		\node[leftnode,on chain] (f5) [label=left: $a$] {};
		\end{scope}
		\begin{scope}[xshift=2.5cm,start chain=going below,node distance=3mm]
		\foreach \i in {1,2,...,4}
			\node[rightnode,on chain] (s\i) [label=right: $b_\i$] {};
		\node[dummynode,on chain] (s5) [label=right: $d_{a_4}$] {};
		\end{scope}
		\node [blue,fit=(f1) (f5),label=above:$A$] {};
		\node [green,fit=(s1) (s4),label=above:$B$] {};
		\draw[red,-] (f1) -- (s2);
		\draw[red,-] (f2) -- (s1) node[above=2mm,midway,red]{$M_{t-1}$};
		\draw[-] (f3) -- (s4);
		\draw[-] (f4) -- (s5);
		\draw[-] (f5) -- (s3);
	\end{tikzpicture}
	\caption{$\mu(A_t,B \setminus \{b\},M_{t-1})$}
	\label{fig:1a}
	\end{subfigure}
	\begin{subfigure}[t]{0.29\linewidth}
	\centering
	\begin{tikzpicture}[bipartite]
		\begin{scope}[start chain=going below,node distance=3mm]
		\foreach \i in {1,2,...,4}
			\node[leftnode,on chain] (f\i) [label=left: $a_\i$] {};
		\end{scope}
		\begin{scope}[xshift=2.5cm,yshift=0.35cm,start chain=going below,node distance=3mm]
		\foreach \i in {1,2,...,4}
			\node[rightnode,on chain] (s\i) [label=right: $b_\i$] {};
		\node[rightnode,on chain] (s5) [label=right: $b$] {};
		\end{scope}
		\node [blue,fit=(f1) (f4),label=above:$A$] {};
		\node [green,fit=(s1) (s5),label=above:$B$] {};

		\draw[red,-] (f1) -- (s2);
		\draw[red,-] (f2) -- (s1) node[above=2mm,midway,red]{$M_{t-1}$};
		\draw[-] (f3) -- (s4);
		\draw[-] (f4) -- (s5);
	\end{tikzpicture}
	\caption{$\mu(A_{t-1},B,M_{t-1})$}
	\label{fig:1b}
	\end{subfigure}
	\begin{subfigure}[t]{0.38\linewidth}
	\centering
	\begin{tikzpicture}[bipartite]
	\node[leftnode,xshift=-1.5cm,yshift=-15mm] (a) [label=above: $s_1$] {};
		\begin{scope}[start chain=going below,node distance=5mm]
		\foreach \i in {1,2,...,4}
			\node[leftnode,on chain] (f\i) [label=above: $a_\i$] {};
		\node[leftnode,on chain] (f5) [label=above: $a$] {};
		\end{scope}
		\begin{scope}[xshift=2cm,start chain=going below,node distance=5mm]
		\foreach \i in {1,2,...,4}
			\node[rightnode,on chain] (s\i) [label=above: $b_\i$] {};
		\node[dummynode,on chain] (s5) [label=above: $d_{a_4}$] {};
		\end{scope}
		\node[rightnode,xshift=3.5cm,yshift=-15mm] (b) [label=above: $t_1$] {};
		\draw (f1) -- (s2);
		\draw (f2) -- (s1);
		\draw (f3) -- (s4);
		\draw (f4) -- (s5);
		\draw (f5) -- (s3);
		\foreach \i in {1,2,...,4}
			\draw (a) -- (f\i);
		\draw (a) -- (f5);
		\foreach \i in {1,2,4}
			\draw (s\i) -- (b);
		\draw (s3) -- (b);
		\draw (s5) -- (b);
	\end{tikzpicture}
	\caption{flow $f_1^*$ over flow-network $N_1$\\ ($0$ flow edges excluded)}
	\label{fig:1c}
	\end{subfigure}
	\caption{Generating flows from matchings (unmatched dummy nodes excluded)}
\end{figure*}
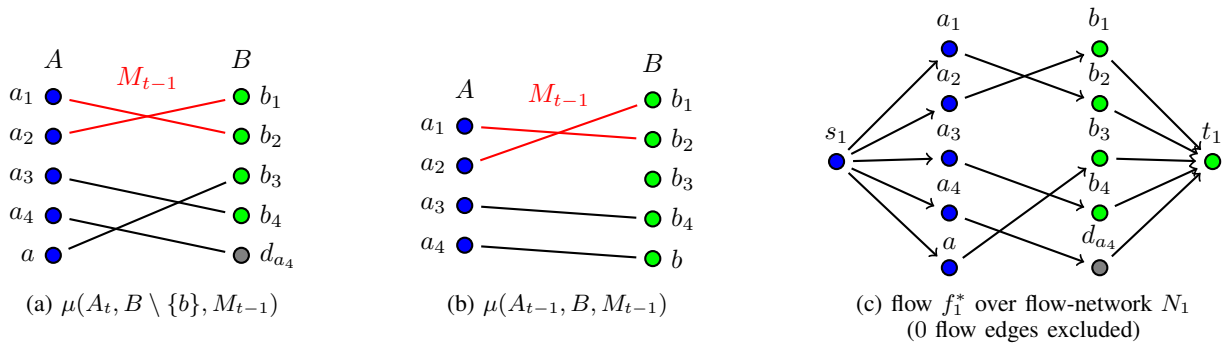

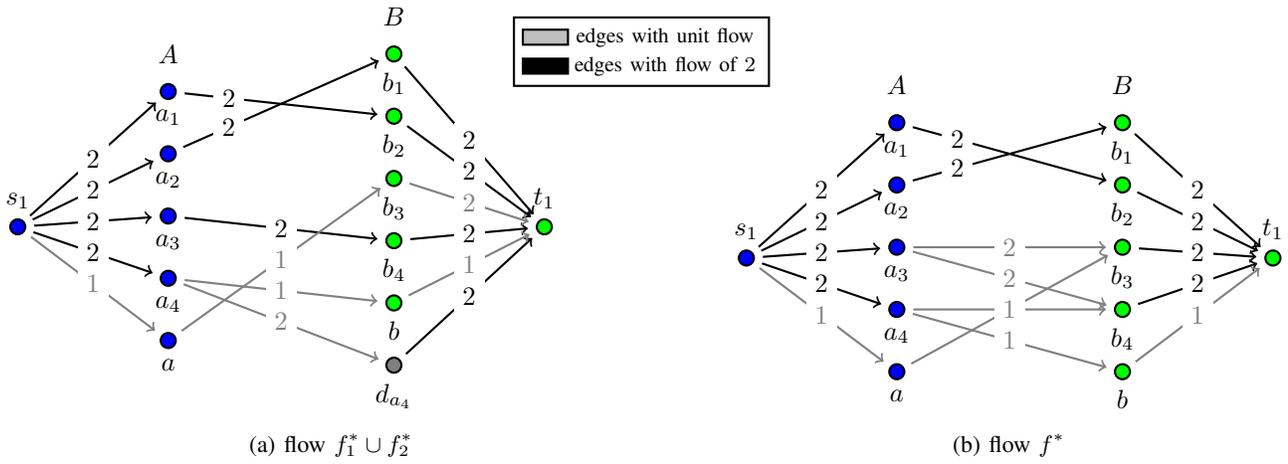
\begin{figure*}[p]
	\begin{subfigure}[b]{0.49\linewidth}
	\centering	
	\begin{tikzpicture}[bipartite]
\begin{customlegend}[
legend entries={ 
edges with unit flow,
edges with flow of $2$
},
legend style={at={(8,1)},font=\footnotesize}] 
    \addlegendimage{area legend,fill=lightgray}
    \addlegendimage{area legend,fill=black}
    \end{customlegend}
	\node[leftnode,xshift=-2cm,yshift=-18mm] (a) [label=above: $s_1$] {};
		\begin{scope}[start chain=going below,node distance=6mm]
		\foreach \i in {1,2,...,4}
			\node[leftnode,on chain] (f\i) [label=below: $a_\i$] {};
		\node[leftnode,on chain] (f5) [label=below: $a$] {};
		\end{scope}
		\begin{scope}[xshift=3cm,yshift=5mm,start chain=going below,node distance=6mm]
		\foreach \i in {1,2,...,4}
			\node[rightnode,on chain] (s\i) [label=below: $b_\i$] {};
		\node[rightnode,on chain] (s5) [label=below: $b$] {};
		\node[dummynode,on chain] (s6) [label=below: $d_{a_4}$] {};
		\end{scope}
		\node[rightnode,xshift=5cm,yshift=-18mm] (b) [label=above: $t_1$] {};
		\node [blue,fit=(f1) (f5),label=above:$A$] {};
		\node [green,fit=(s1) (s4),label=above:$B$] {};
		\draw[] (f1) -- (s2) node[centerlabel,pos=0.25]{$2$};
		\draw[] (f2) -- (s1) node[centerlabel,pos=0.25]{$2$};
		\draw[] (f3) -- (s4) node[centerlabel,pos=0.5]{$2$};
		\draw[gray] (f4) -- (s5) node[centerlabel,pos=0.5]{$1$};
		\draw[gray] (f4) -- (s6) node[centerlabel,pos=0.5]{$2$};
		\draw[gray] (f5) -- (s3) node[centerlabel,pos=0.5]{$1$};
		\foreach \i in {1,2,...,4}
			\draw[] (a) -- (f\i) node[centerlabel,pos=0.5]{$2$};
		\draw[gray] (a) -- (f5) node[centerlabel,pos=0.5]{$1$};
		\foreach \i in {1,2,4,6}
			\draw[] (s\i) -- (b) node[centerlabel,pos=0.5]{$2$};
		\draw[gray] (s3) -- (b) node[centerlabel,pos=0.5]{$2$};
		\draw[gray] (s5) -- (b) node[centerlabel,pos=0.5]{$1$};
	\end{tikzpicture}
	\caption{flow $f_1^* \cup f_2^*$}
	\label{fig:2a}
	\end{subfigure}
	\begin{subfigure}[b]{0.49\linewidth}
	\centering	
	\begin{tikzpicture}[bipartite]
	\node[leftnode,xshift=-2cm,yshift=-18mm] (a) [label=above: $s_1$] {};
		\begin{scope}[start chain=going below,node distance=6mm]
		\foreach \i in {1,2,...,4}
			\node[leftnode,on chain] (f\i) [label=below: $a_\i$] {};
		\node[leftnode,on chain] (f5) [label=below: $a$] {};
		\end{scope}
		\begin{scope}[xshift=3cm,start chain=going below,node distance=6mm]
		\foreach \i in {1,2,...,4}
			\node[rightnode,on chain] (s\i) [label=below: $b_\i$] {};
		\node[rightnode,on chain] (s5) [label=below: $b$] {};
		\end{scope}
		\node[rightnode,xshift=5cm,yshift=-18mm] (b) [label=above: $t_1$] {};
		\node [blue,fit=(f1) (f5),label=above:$A$] {};
		\node [green,fit=(s1) (s4),label=above:$B$] {};
		\draw (f1) -- (s2) node[centerlabel,pos=0.25]{$2$};
		\draw (f2) -- (s1) node[centerlabel,pos=0.25]{$2$};
		\draw[gray] (f3) -- (s3) node[centerlabel,pos=0.5]{$2$};
		\draw[gray] (f3) -- (s4) node[centerlabel,pos=0.5]{$2$};
		\draw[gray] (f4) -- (s5) node[centerlabel,pos=0.5]{$1$};
		\draw[gray] (f4) -- (s4) node[centerlabel,pos=0.5]{$2$};
		\draw[gray] (f5) -- (s3) node[centerlabel,pos=0.5]{$1$};
		\foreach \i in {1,2,...,4}
			\draw[] (a) -- (f\i) node[centerlabel,pos=0.5]{$2$};
		\draw[gray] (a) -- (f5) node[centerlabel,pos=0.5]{$1$};
		\foreach \i in {1,2,4}
			\draw[] (s\i) -- (b) node[centerlabel,pos=0.5]{$2$};
		\draw (s3) -- (b) node[centerlabel,pos=0.5]{$2$};
		\draw[gray] (s5) -- (b) node[centerlabel,pos=0.5]{$1$};
	\end{tikzpicture}
	\caption{flow $f^*$}
	\label{fig:2b}
	\end{subfigure}
	\caption{Flow network $N_{12}$ with capacity on edges (edges without flow are excluded)}
\end{figure*}

\begin{figure*}[p]
	\centering	
	\begin{tikzpicture}[bipartite]
\begin{customlegend}[
legend entries={ 
edges added to $F_1$,
edges added to $F_2$,
edges added to $F_1$ and $F_2$
},
legend style={at={(11,0)},font=\footnotesize}] 
    \addlegendimage{area legend,fill=olive}
    \addlegendimage{area legend,fill=teal}
    \addlegendimage{area legend,fill=purple}
    \end{customlegend}
	\node[leftnode,xshift=-2cm,yshift=-15mm] (a) [label=above: $s_1$] {};
		\begin{scope}[start chain=going below,node distance=5mm]
		\foreach \i in {1,2,...,4}
			\node[leftnode,on chain] (f\i) [label=below: $a_\i$] {};
		\node[leftnode,on chain] (f5) [label=below: $a$] {};
		\end{scope}
		\begin{scope}[xshift=3cm,start chain=going below,node distance=5mm]
		\foreach \i in {1,2,...,4}
			\node[rightnode,on chain] (s\i) [label=below: $b_\i$] {};
		\node[rightnode,on chain] (s5) [label=below: $b$] {};
		\end{scope}
		\node[rightnode,xshift=5cm,yshift=-15mm] (b) [label=above: $t_1$] {};
		\node [blue,fit=(f1) (f5),label=above:$A$] {};
		\node [green,fit=(s1) (s4),label=above:$B$] {};
		\draw[purple] (f1) -- (s2) node[centerlabel,pos=0.25]{$2$};
		\draw[purple] (f2) -- (s1) node[centerlabel,pos=0.25]{$2$};
		\draw[teal] (f3) -- (s3) node[centerlabel,pos=0.5]{$2$};
		\draw[olive] (f3) -- (s4) node[centerlabel,pos=0.5]{$2$};
		\draw[olive] (f4) -- (s5) node[centerlabel,pos=0.5]{$1$};
		\draw[teal] (f4) -- (s4) node[centerlabel,pos=0.5]{$2$};
		\draw[olive] (f5) -- (s3) node[centerlabel,pos=0.5]{$1$};
		\foreach \i in {1,2,...,4}
			\draw[purple] (a) -- (f\i) node[centerlabel,pos=0.5]{$2$};
		\draw[olive] (a) -- (f5) node[centerlabel,pos=0.5]{$1$};
		\foreach \i in {1,2,...,4}
			\draw[purple] (s\i) -- (b) node[centerlabel,pos=0.5]{$2$};
		\draw[olive] (s5) -- (b) node[centerlabel,pos=0.5]{$1$};
	\end{tikzpicture}
	\caption{decomposing flow $f^*$ into sets $F_1$ and $F_2$}
	\label{fig:3}
\end{figure*}
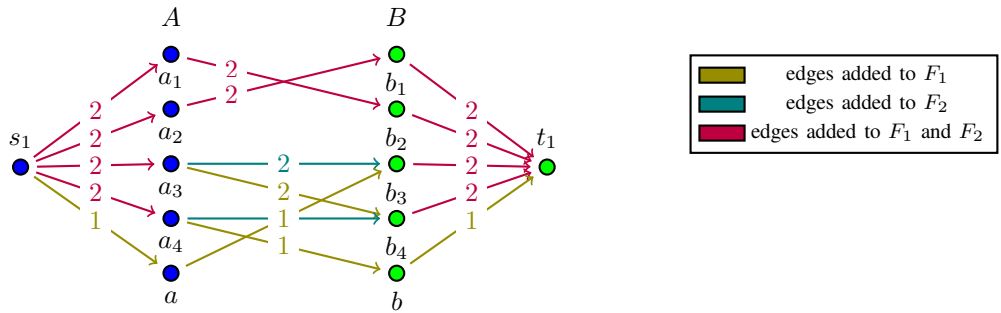

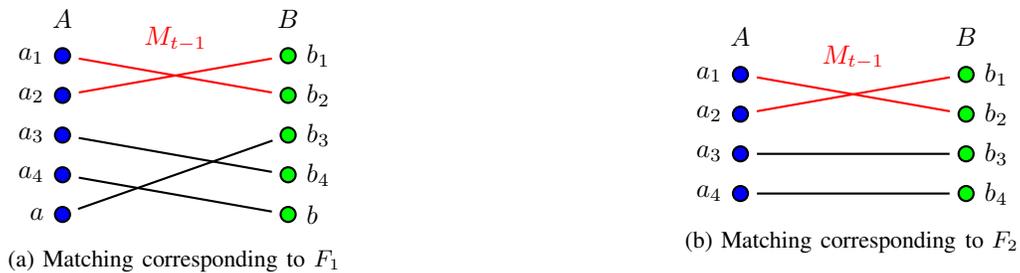
\begin{figure*}[p]
	\begin{subfigure}[c]{0.49\linewidth}
	\centering
	\begin{tikzpicture}[bipartite]
		\begin{scope}[start chain=going below,node distance=3mm]
		\foreach \i in {1,2,...,4}
			\node[leftnode,on chain] (f\i) [label=left: $a_\i$] {};
		\node[leftnode,on chain] (f5) [label=left: $a$] {};
		\end{scope}
		\begin{scope}[xshift=3cm,start chain=going below,node distance=3mm]
		\foreach \i in {1,2,...,4}
			\node[rightnode,on chain] (s\i) [label=right: $b_\i$] {};
		\node[rightnode,on chain] (s5) [label=right: $b$] {};
		\end{scope}
		\node [blue,fit=(f1) (f5),label=above:$A$] {};
		\node [green,fit=(s1) (s4),label=above:$B$] {};
		\draw[red,-] (f1) -- (s2);
		\draw[red,-] (f2) -- (s1) node[above=2mm,midway,red]{$M_{t-1}$};
		\draw[-] (f3) -- (s4);
		\draw[-] (f4) -- (s5);
		\draw[-] (f5) -- (s3);
	\end{tikzpicture}
	\caption{Matching corresponding to $F_1$}
	\label{fig:4a}
	\end{subfigure}
\begin{subfigure}[c]{0.49\linewidth}
	\centering
	\begin{tikzpicture}[bipartite]
		\begin{scope}[start chain=going below,node distance=3mm]
		\foreach \i in {1,2,...,4}
			\node[leftnode,on chain] (f\i) [label=left: $a_\i$] {};
		\end{scope}
		\begin{scope}[xshift=3cm,start chain=going below,node distance=3mm]
		\foreach \i in {1,2,...,4}
			\node[rightnode,on chain] (s\i) [label=right: $b_\i$] {};
		\end{scope}
		\node [blue,fit=(f1) (f4),label=above:$A$] {};
		\node [green,fit=(s1) (s4),label=above:$B$] {};

		\draw[red,-] (f1) -- (s2);
		\draw[red,-] (f2) -- (s1) node[above=2mm,midway,red]{$M_{t-1}$};
		\draw[-] (f3) -- (s3);
		\draw[-] (f4) -- (s4);
	\end{tikzpicture}
	\caption{Matching corresponding to $F_2$}
	\label{fig:4b}
	\end{subfigure}
	\caption{Matchings constructed from $F_1$ and $F_2$ (unmatched dummy nodes excluded)}
\end{figure*}

    In the following, we decompose the edges in $f^*$ into $2$ sets $F_1$ and $F_2$, and show that the $F_1$ corresponds to a valid matching between $A_t$ and $B$ with the condition that all edges in $L_t$ must be included, while $F_2$ corresponds to a valid matching between $A_{t-1}$ and $B \setminus \{b\}$ under the same condition that the edges in $L_t$ are included in the matching. The decomposition is as follows:
    \begin{enumerate}
    \item by definition of $f^*$, every edge to $A_t$ from $s_1$ is saturated. Therefore, edges to $A_t \setminus \{a\}$ are split into two each having unit capacity, unit flow and weight unchanged as before, and given to $F_1$ and $F_2$. In Figure \ref{fig:3}, these correspond to the edges $(s_1,a_1),(s_1,a_2),(s_1,a_3)$ and $(s_1,a_4)$. The edge $(s_1,a)$ is also given to $F_1$.
    \item Consider those edges in $f^*$ between $A_t$ and $B \cup D$ (all such edges have capacity at-most $2$) that are saturated with $2$ units of flow. In Figure \ref{fig:3}, these correspond to the edges $(a_1,b_2)$ and $(a_2,b_1)$. Every edge in $L_t$ satisfies this criterion. Each such edge is divided into two identical $1$-capacity edges with unchanged weight and unit flow through them, and one is added to $F_1$ while the other is added to $F_2$. These edges are then removed from $f^*$. The rest of the edges between $A_t$ and $B \cup D$ in $f^*$ must have unit flow through them.
    \item At the end of Step 1, every remaining edge in $f^*$ between $A_t$ and $B \cup D$ has unit flow through it. Starting from any vertex in $A_t \cup B \cup D$ on which is incident a \textbf{single} edge with unit flow through it ($a$ and $b$ are valid such starting points), form a path alternating between vertices in $A_t$ and $B \cup D$ using the edges remaining after Step 1 in $f^*$. Edges in this path are placed alternately into $F_1$ and $F_2$ in such a way that the edge involving $a, b$ or $d_a$ (dummy node corresponding to $a$) is put into $F_1$. Referring back to Figure \ref{fig:3}, starting from $a$, we follow the path ${\color{olive} (a,b_3)} \to {\color{teal}(b_3,a_3)} \to {\color{olive}(a_3,b_4)} \to {\color{teal}(b_4,a_4)} \to {\color{olive}(a_4,b)}$ and add odd edges to $F_1$ and even edges to $F_2$. It is always possible for edges involving $a$, $b$ or $d_a$ be put into $F_1$, since any path starting in $A_t$ and ending in $B \cup D$ or vice-versa would have to be of odd parity (if $(a,v_1)$ is placed in $F_1$ for some $v_1$, then $(v_2,b)$ for some $v_2$ would have to be placed in $F_1$ without exception). On the other hand, if a unit-flow edge is incident on $d_a$:
    \begin{enumerate}
        \item it must be connected to $a$ (by definition of dummy edge).
        \item only one such unit-flow edge exists (since no edge can exist from $d_a$ to any other vertex).
    \end{enumerate}
    Hence if $a$ is placed in $F_1$ and a unit flow edge to $d_a$ exists, the path must have a single edge $(a,d_a)$. The set of edges along this path is removed from $f^*$. Step 2 is repeated until no vertex remains in $A_t \cup B \cup D$ onto which a single unit-flow edge in $f^*$ is incident.
    \item Since no vertex in $A_t \cup B \cup D$ with a single unit-flow edge incident on it in $f^*$ remains, the remaining unit-flow edges in $f^*$ must form cycles that alternate between $A_t$ and $B \cup D$. Moreover, such cycles are guaranteed to be simple - for every every vertex $a' \in A_t$ the incoming flow along the edge to $s_1$ cannot exceed $2$ units and hence $a'$ can only have at-most $2$ unit-flow edges incident on it. Similarly, for every vertex $b' \in B \cup D$ the outgoing flow along the edge to $t_1$ cannot exceed $2$ units, implying that it can cannot have more than $2$ unit-flow edges incident on it. Such cycles are also guaranteed to have an even number of edges which follows by a simple parity argument. Each edge in such a cycle is alternately assigned to $F_1$ and $F_2$ with arbitrary assignment to the starting vertex.
    \item The final set of remaining edges $E_{t_1}$ are those from $B \cup D$ to $t_1$. Every edge having $2$ units of flow through it in $f^*$ is split into unit capacity edges having unchanged weight and unit-flow through them. A subset $E_1$ of these edges is chosen such that for $F_1$, the flow input into every vertex in $B \cup D$ is matched by the flow output due to the corresponding edge in $E_1$. The set $E_1$ is added to $F_1$. The remaining edges $E_{t_1} \setminus E_1$ match the flow input for every vertex in $B \cup D$ in $F_2$. This is because $f^*$ is a valid flow, and hence the flow input at every vertex in $B \cup D$ due to $F_1 \cup F_2$, is exactly matched by the corresponding edge in $E_{t_1}$.
    \end{enumerate}
    
    By this construction, the set $F_1$ represents a valid flow over a network with source $s_1$, sink $t_1$, left nodes $A_t$ and right nodes $B \cup D$ and every edge in $L_t$ has unit flow through it. On the other hand $F_2$ represents a valid flow over a network with source $s_1$, sink $t_1$, left nodes $A_{t} \setminus \{a\}$ and right nodes $(B \setminus \{b\}) \cup (D \setminus \{d_a\})$ and once again, every edge in $L_t$ has unit flow through it. Thus, the sum of weights of these two flows, which is equal to $H(f^*)$ must be upper bounded by the maximum flows over the respective networks (conditioned on the fact that edges in $L_t$ must all have unit flow through them). Thus,
    \begin{align}
        H(f^*) &\le W(A_t, B, L_t) + W(A_t \setminus \{a\}, B \setminus \{b\}, L_t), \nonumber \\
        &\overset{(i)}{=} W(A_t, B, L_t) + W(A_{t-1}, B \setminus \{b\}, L_{t-1}), \label{eq:004}
    \end{align}
    where $(i)$ follows from the fact that $a$ is the node that arrives at time $t$, and also from the fact that no node is locked in at time $t$ and hence $L_t = L_{t-1}$. Thus from (\ref{eq:003}) and (\ref{eq:004}), it follows that
    \begin{equation*}
        \rho_{t-1}(b) \le \rho_t(b).
    \end{equation*}

\end{proof}

\section{Proof of Theorem \ref{theorem:12apx}}
To prove Theorem \ref{theorem:12apx}, we make use of the well known online submodular maximization problem \cite{Fisher1978} (defined next)  for which the online-greedy algorithm is guaranteed to be $2$-competitive.
\begin{problem}\label{prob:stand}[Standard Online Submodular Maximization Problem]
Elements $r\in R$ arrive online in any arbitrary order, and a set of bins $B$ are available before hand. Each $r$ on its arrival must be immediately and irrevocably allocated to some bin of $B$. Given a monotone and submodular set-function $Y(S) : 2^{R \times B} \to \mathbb{R}^+$, the objective is to find a valid allocation $S$ of resources $R$ to bins $B$ that maximizes $Y(S)$. 
\end{problem}
The usual greedy algorithm \cite{Fisher1978} for the standard online submodular maximization problem assigns each new resource to the bin that provides the large incremental gain in valuation, and will be referred to as $\mathsf{GREEDY-ON}$ for the rest of the appendix. Recall that $\mathsf{GREEDY-ON}$ breaks ties between bins arbitrarily if there is more than one bin with the same incremental valuation. We will use a particular tie-breaking rule for our purposes.
\begin{proof}[Proof of Theorem \ref{theorem:12apx}]

To prove Theorem \ref{theorem:12apx}, we consider an arbitrary instance of Problem \ref{RA:locking}, $I$ with valuation function $Z(\cdot)$ and reduce it to $I_{SM}$ - an instance of the standard online submodular maximization with valuation $Y(\cdot)$ (Problem \ref{prob:stand}) which satisfies the following properties:
\begin{enumerate}
\item The valuation of the offline optimal solution to $I$ is upper bounded by the valuation of the offline optimal solution to $I_{SM}$ (shown in Lemma \ref{lemma:L-RA}). In other words, we show that $Y(\Omega_{SM}) \ge Z(\Omega)$, where $\Omega_{SM}$ is the offline optimal solution to $I_{SM}$ and $\Omega$ is the offline optimal solution to Problem \ref{RA:locking}.
\item The valuation of $G$, the solution generated by the Algorithm \ref{alg:on-greedy} on $I$ is equal to the valuation of $G_{SM}$, the solution generated by $\mathsf{GREEDY-ON}$ on $I_{SM}$ (with a particular tie-breaking rule). In other words, $Z(G) = Y(G_{SM})$. This is shown in Lemma \ref{lemma:greedy-equal}. 
\end{enumerate}
Since $\mathsf{GREEDY-ON}$ is known to be $2$-competitive for any instance of the standard online submodular maximization problem \cite{Fisher1978}, in particular $I_{SM}$, we have that $\frac12 Y(\Omega_{SM}) \le Y(G_{SM})$. From the relation between $Y(G_{SM})$ and $Z(G)$ and between $Z(\Omega)$ and $Y(\Omega_{SM})$ established in Lemma \ref{lemma:L-RA} and Lemma \ref{lemma:greedy-equal} as described above, it follows that Algorithm \ref{alg:on-greedy} is $2$-competitive for Problem \ref{RA:locking}.
\end{proof}



In the rest of this section, we prove Lemma \ref{lemma:L-RA} and Lemma \ref{lemma:greedy-equal}. Towards that end, we use the following notation in the context of the standard online submodular maximization problem (Problem \ref{prob:stand}):
\begin{enumerate}
\item As in Problem \ref{RA:locking}, we use $T$ to refer to the time horizon and $N$ as the total number of resources that arrive till time $T$.
\item For some $k \le N$, a \textit{partial solution} $S_1$ is a set of tuples $\{(r_i,b_i), i=1,2,\dots,k\}$ with $r_i \ne r_j$ for $i \ne j$. A \textit{feasible solution} is a partial solution of cardinality $N$.
\item Given $S \subseteq R \times B$, we define $\mu(r,b|S)$ as $Y(S \cup \{(r,b)\}) - Y(S)$, the increment in valuation of $S$ when the resource-bin tuple $(r,b)$ is added to $S$. Given any feasible solution $S = \{(r_i,b_i),\ i=1,2,\dots,N\}$ to the standard online submodular maximization problem, with $S^j = \{(r_i,b_i),\ i=1,2,\dots,j\}$, $Y(S)$ can be decomposed as:
\begin{equation} \label{eq:RA:decomp}
Y(S) = \sum_{i=1}^N \mu(r_i,b_i|S^{i-1}).
\end{equation}
\end{enumerate}

\noindent We now define $I_{SM}$, the instance of standard online submodular maximization which is constructed from $I$, the arbitrary instance of Problem \ref{RA:locking} we started out with. The set of bins and resources in $I_{SM}$ are identical to that of $I$. The order of arrival of resources in $I_{SM}$ is also identical to that in $I$. The valuation function of $I_{SM}$ is defined in terms of its increments $\mu(r,b|S)$ as:
\begin{equation} \label{eq:newvals}
    \mu(r,b|S) =
    \begin{cases}
    \rho(r,b|S), \qquad &\forall r : T_r \le T_b, \\
    0, &\forall r : T_r > T_b,
    \end{cases}
\end{equation}
where recall that $\rho(r,b|S)$ is the incremental valuation of adding resource $r$ to bin $b$ (in the solution $S$) in the instance $I$. The valuation function of $I_{SM}$ is closely tied to that of $I$ - a bin in $I_{SM}$ assigns $0$ value to all resources that arrive after its counterpart in $I$ locks. As an example, consider the case when $I$ is the general AQI problem (for which $Z(S)$ is given in (\ref{eq:Z(S)-val}) and (\ref{eq:incremental-val})). To illustrate the newly constructed instance $I_{SM}$ of standard online submodular maximization, consider the simple case of some packet $p$ having only one subpacket $s$ which arrives at time $t_0$. Thus, $t_o= T_r$ and $t=T_b$. From (\ref{eq:newvals}), for every time slot $b_{t}$ where $t \ge t_0$, the incremental valuation is defined as:
\begin{equation} \label{eq:example}
\mu (s, b_t | S ) = D_p (1) - \Delta g (S_{b_{t}}) -C_p (t - t_0).
\end{equation}
For this example, in the instance $I$, scheduling subpacket $s$ in slot $b_t$ at any time before $t$ accrues the incremental valuation $D_p (1) - \Delta g (S_{b_{t}}) -C_p (t - t_0)$. But beyond time $t$, subpacket $s$ is not allowed to be scheduled in the slot $b_t$ because it is locked. It is important to note that this notion of "locking" in $I$ is no longer present in $I_{SM}$ - the subpacket $s$ can be scheduled in the slot $b_t$ even after time $t$ and the increment in valuation is fixed as the quantity in (\ref{eq:example}) for posterity. Similarly, for every time slot $b_t$ where $t < t_0$, $\mu (s, b_t | S )$ is fixed as $0$ which and is unaffected by the locking of any bins in $I$.

With $\mathsf{GREEDY-ON}$, each resource $r_i$ is allocated to the bin $b$ that gives the best local improvement to the partial solution maintained till that time. Using $G_{SM}^{i-1}$ to denote the partial solution output by $\mathsf{GREEDY-ON}$ till time $T_{r_i}$ (arrival time of $r_i$) and $\mu(r,b|S)$ to denote $Z(S \cup \{(r,b)\}) - Z(S)$, $\mathsf{GREEDY-ON}$ allocates resource $r_i$ to the bin $\hat{b}_i$ that satisfies:
\begin{align} \label{eq:on-greedy-update}
\hat{b}_i &= \underset{b \in B}{\text{argmax}} \mu(r_i,b_i|G_{SM}^{i-1}),\\
& \text{where } \forall j, G_{SM}^j = G_{SM}^{j-1} \cup \{(r_j,\hat{b}_j)\} \ \left(\text{tie broken arbitrarily}\right)
\end{align}
$\mathsf{GREEDY-ON}$ has an arbitrary tie-breaking rule.
\begin{lemma} \label{lemma:L-RA}
Denoting $\Omega$ as the optimal-offline solution to $I$, and $\Omega_{SM}$ as the optimal-offline solution to $I_{SM}$, we have $Z(\Omega) \le Y(\Omega_{SM})$.
\end{lemma}
\begin{proof}
Recall that the order of arrival of resources is specified by $R_{on} = (r_1,r_2,\dots,r_N)$. We use $\Omega = \{(r_i,b_i^*),\ i=1,2,\dots,N\}$ to denote the optimal-offline solution to Problem \ref{RA:locking} and $\Omega^j$ to denote $\{(r_i,b_i^*),\ i=1,2,\dots,j\}$.
	In any feasible solution to Problem \ref{RA:locking}, notably $\Omega$, a resource $r$ cannot be allocated to a bin $b \in B$ that has already locked before it arrives - any resource $r$ can only be allocated to a bin belonging to the set $\{b \in B : T_b \ge T_r\}$. Therefore, $b_i^*$ must belong to the set $\{b \in B : T_b \ge T_{r_i}\}$, and hence:
\begin{equation} \label{eq:ordering}
T_{r_i} \le T_{b_i^*}
\end{equation}
Using the definition of $\mu (r,b | S)$ in (\ref{eq:newvals}), it follows from (\ref{eq:ordering}) that:
\begin{equation} \label{eq:singleiter}
\mu (r_i,b_i^* | \Omega^{i-1}) = \rho (r_i,b_i^* | \Omega^{i-1}).
\end{equation}
Summing both sides of (\ref{eq:singleiter}) over $i$,
     \begin{align}
        Z(\Omega) &= \sum_{i=1}^N \mu(r_i,b_i^* | \Omega^{i-1}), \\
        &\overset{(i)}{=} Y(\Omega), \label{eq:ZY}
	\end{align}
where $(i)$ follows from (\ref{eq:RA:decomp}). We also have that 
\begin{equation} \label{eq:laststep}
Y(\Omega) \le Y(\Omega_{SM})
\end{equation}
This is because $\Omega_{SM}$ is the optimal solution to $I_{SM}$, and $\Omega$ is a feasible solution to $I_{SM}$ as it satisfies the condition $r_i \ne r_j$ for $i \ne j$ (any feasible solution to Problem \ref{RA:locking} must satisfy this condition). Putting (\ref{eq:laststep}) together with (\ref{eq:ZY}) completes the proof.
\end{proof}


\noindent For the purpose of Lemma \ref{lemma:greedy-equal}, we now introduce a tie-breaking rule for $\mathsf{GREEDY-ON}$. In Lemma \ref{lemma:greedy-equal}, we show that $\mathsf{GREEDY-ON}$ with such a tie-breaking rule generates a solution $G_{SM}$ with valuation $Y(G_{SM})$ equal to $Z(G)$ where $G$ is the solution generated by Algorithm \ref{alg:on-greedy}.

\paragraph*{Tie-breaking rule for $\mathsf{GREEDY-ON}$, $\mathsf{RULE}$}  \label{tie-breaking-rule}
Denote the partial solution generated by $\mathsf{GREEDY-ON}$ after the first $i-1$ resources arrive as $G_{SM}^{i-1} = \{(r_j,\hat{b}_j),\ j=1,2,\dots,i-1\}$. 
At time $T_{r_i}$, resource $r_i$ arrives:

\begin{enumerate}
\item If $\max_{b \in B} \mu(r,b|G_{SM}^{i-1}) > 0$ and there is a tie, break the tie arbitrarily (allocate $r_i$ to any of the tied $b$ chosen arbitrarily). In this case, the bin allocated to after tie-breaking would be such that $T_{r_i} \le T_b$, since from (\ref{eq:newvals}), every bin such that $T_{r_i} \ge T_b$ has $\mu(r_i,b|G_{SM}^{i-1}) = 0$ and we are guaranteed that atleast one bin exists such that $\max_{b \in B} \mu(r_i,b|G_{SM}^{i-1}) > 0$.
\item If $\max_{b \in B} \mu(r_i,b|G_{SM}^{i-1}) = 0$ and there is a tie, the tie is broken arbitrarily by allocating $r_i$ to any tied bin $b$ that satisfies the condition $T_{r_i} \le T_b$. This set cannot be empty, as we are guaranteed that $b_d$ always satisfies this condition, since $T_{r_i} \le T_{b_d} = \infty$ and $\mu(r_i,b_d|G_{SM}^{i-1}) = \rho(r_i,b_d | G^{i-1}_{SM}) = 0$.
\end{enumerate}
In both cases, $\mathsf{GREEDY-ON}$ allocates $r_i$ to a bin $b : T_{r_i} \le T_b$. Thus, for $\mathsf{GREEDY-ON}$ with $\mathsf{RULE}$, we have:
\begin{align}
\hat{b}_i &= \underset{b \in B}{\text{argmax}}  \left\{\mu(r_i,b|G_{SM}^{i-1})\right\}, \nonumber \\
&= \underset{b \in B : T_{r_i} \le T_b}{\text{argmax}}  \left\{\mu(r_i,b|G_{SM}^{i-1})\right\}, \nonumber \\
&\overset{(i)}{=} \underset{b \in B : T_{r_i} \le T_b}{\text{argmax}}  \left\{\rho(r_i,b|G_{SM}^{i-1})\right\}, \label{eq:tie-inc}
\end{align}
where $(i)$ follows from the definition of $\mu(r,b|S)$ in (\ref{eq:newvals}). From (\ref{eq:newvals}) it also follows that since $T_{r_i} \le T_{\hat{b}_i}$, 
$\mu(r_i,\hat{b}_i | G^{i-1}_{SM}) = \rho(r_i,\hat{b}_i | G^{i-1}_{SM})$. Therefore, 
\begin{align} \label{eq:tie-vals}
Y(G_{SM}) &= \sum_{i=1}^N \mu(r_i,\hat{b}_i | G_{SM}^{i-1})= \sum_{i=1}^N \rho(r_i,\hat{b}_i | G^{i-1}_{SM}). 
\end{align}

\begin{lemma} \label{lemma:greedy-equal}
Denote the online solution generated by Algorithm \ref{alg:on-greedy} on the instance $I$ as $G$. Denoting the online solution generated by $\mathsf{GREEDY-ON}$ with $\mathsf{RULE}$ on  $I_{SM}$ as $G_{SM}$, we have that $Z(G) =Y(G_{SM})$.
\end{lemma}
\begin{proof}

Let Algorithm \ref{alg:on-greedy} generate a solution $G = \{(r_i,b_i),\ i =1,2,\dots,N\}$ and let $G^j$ denote $\{ (r_i,b_i),\ i =1,2,\dots,j\}$. By definition of Algorithm \ref{alg:on-greedy}, for every $i$,
\begin{align}
b_i &= \underset{b \in B_{t}}{\text{argmax}} \left\{\rho(r_i,b|G^{i-1})\right\}, \quad \text{where } t=T_{r_i}. \label{eq:bi-alg}
\end{align}
Note that the set $B_{t}$ is the set of unlocked bins at time $t$ and is hence equal to $\{b \in B: t \le T_b\}$. Therefore, at time $t=T_{r_i}$, $B_{t} = \{b \in B: T_{r_i} \le T_b\}$. Substituting this expression for $B_{t_i}$ in (\ref{eq:bi-alg}), we get:
\begin{equation} \label{eq:bi-alg-final}
b_i = \underset{b \in B : T_{r_i} \le T_b}{\text{argmax}} \left\{\rho(r_i,b|G^{i-1})\right\}.
\end{equation}

\noindent Let $\mathsf{GREEDY-ON}$ with $\mathsf{RULE}$ generate solution $G_{SM}  = \{(r_i,\hat{b}_i),\ i =1,2,\dots,N\}$ and let $G_{SM}^j$ denote the partial solution $\{(r_i,\hat{b}_i),\ i =1,2,\dots,j\}$. From (\ref{eq:tie-inc}) it follows that resource $r_i$ is allocated to the bin $\hat{b}_i$ such that
\begin{equation} \label{eq:bi-on-gr-final}
\hat{b}_i = \underset{b \in B : T_{r_i} \le T_b}{\text{argmax}} \left\{\rho(r_i,b|G_{SM}^{i-1})\right\}.
\end{equation}
Thus, using (\ref{eq:bi-alg-final}) and (\ref{eq:bi-on-gr-final}), we use an inductive argument to show that:
\begin{equation} \label{eq:incrementally-equal} 
\forall i, b_i = \hat{b}_i \text{ and } G^{i}_{SM} = G^i,
\end{equation}
\begin{enumerate}
\item for $i=1$, $G_{SM}^0 = G^0 = \phi$. Therefore $\hat{b}_i = b_i = \underset{b \in B : T_{r_i} \le T_b}{\text{argmax}} \left\{\rho(r_i,b|\phi)\right\}$.
\item Assume that the statement is true for $i=j-1$. For $i=j$, we have that $\hat{b}_j = b_j$ since: 
\begin{align}
\hat{b}_j &= \underset{b \in B : T_{r_j} \le T_b}{\text{argmax}} \left\{\rho(r_i,b|G_{SM}^{i-1})\right\} \nonumber \\
&\overset{(i)}{=} \underset{b \in B : T_{r_j} \le T_b}{\text{argmax}} \left\{\rho(r_i,b|G_\perp^{i-1})\right\} = b_j. \label{eq:bi=hatbi}
\end{align}
where in $(i)$ we use the fact that the statement is true for $i=j-1$ to conclude that $G^{j-1} = G_{SM}^{j-1}$. The proof that $G^j = G^j_{SM}$ also follows swiftly, since:
\begin{align*}
G^j &= G^{j-1} \cup \{(r_j,b_j)\}, \\
&\overset{(i)}{=} G_{SM}^{j-1} \cup \{(r_j,\hat{b}_j)\} = G^j_{SM}. 
\end{align*}
where $(i)$ follows from (\ref{eq:bi=hatbi}) and using the fact that $G^{j-1} = G_{SM}^{j-1}$ (since the statement is true for $i=j-1$). This concludes the proof by induction.
\end{enumerate}
Therefore, we have:
\begin{align*}
Z(G) &= \sum_{i=1}^N \rho(r_i,b_i | G^{i-1}), \\
&\overset{(i)}{=} \sum_{i=1}^N \rho(r_i,\hat{b}_i | G_{SM}^{i-1}) \overset{(ii)}{=} Y(G_{SM}),
\end{align*}
where $(i)$ follows from (\ref{eq:incrementally-equal}) and $(ii)$ follows from (\ref{eq:tie-vals}).
\end{proof}

\end{document}